\documentclass[a4paper,10pt]{article}
\usepackage{amsmath,amssymb,mathtools,cite,graphicx,fancyhdr}
\usepackage{bbold,caption,mathpazo,authblk}
\usepackage[dvipsnames]{color}
\usepackage[colorlinks = true, linkcolor = blue, urlcolor  = black, citecolor = magenta]{hyperref}
\usepackage[strings]{underscore}
\graphicspath{{Graphs/}}
\newcommand{\dd}{\text{d}}
\newcommand{\xtilde}{{\raise.17ex\hbox{$\scriptstyle\sim$}}}

\begin{document}
\lhead{The Theory of Optical Black Hole Lasers}
\rhead{Gaona-Reyes, Bermudez}

\title{The Theory of Optical Black Hole Lasers}
\author{Jos\'e L. Gaona-Reyes\footnote{{\it email:} jgaona@fis.cinvestav.mx} }
\author{David Bermudez\footnote{\textit{email:} dbermudez@fis.cinvestav.mx; \textit{web:} \href{http://www.fis.cinvestav.mx/\textasciitilde dbermudez/}{http://www.fis.cinvestav.mx/\xtilde dbermudez}
}}

\affil{\textit{Department of Physics, Cinvestav, A.P. 14-740, 07000 Mexico City, Mexico}}

\date{}
\maketitle

\begin{abstract}
The event horizon of black holes and white holes can be achieved in the context of analogue gravity. It was proven for a sonic case that if these two horizons are close to each other their dynamics resemble a laser, a black hole laser, where the analogue of Hawking radiation is trapped and amplified. Optical analogues are also very successful and a similar system can be achieved there. In this work we develop the theory of optical black hole lasers and prove that the amplification is also possible. Then, we study the optical system by determining the forward propagation of modes, obtaining an approximation for the phase difference which governs the amplification, and performing numerical simulations of the pulse propagation of our system.\\

{\it Keywords:} black hole laser; analogue gravity; Hawking radiation; pulse dynamics; negative frequency.
\end{abstract}

\section{Introduction} \label{intro}
The emerging field of analogue gravity \cite{Barcelo2011,Leonhardt2015pta,Bermudez2016var} has been successful in relating gravity with analogue systems taken from different fields of physics such as water waves \cite{Unruh2007,Rousseaux2010}, Bose-Einstein condensates (BECs) \cite{Barcelo2001}, optical pulses in fibers \cite{Philbin2008}, and more recently, Type-II Weyl Fermions \cite{Volovik2016} and magnetization dynamics \cite{RoldanMolina2016}. Furthermore, advances in both gravitational and analogue systems have been possible through the study of these analogies. Part of their success is due to the interplay between theoretical models and experimental results \cite{Philbin2008,Belgiorno2010,Steinhauer2016}. One of the main topics of interest is the creation of particles at the event horizon of a black hole, i.e., Hawking radiation.

S. Hawking predicted in 1974 that black holes emit particles with a thermal spectrum \cite{Hawking1974} and W. Unruh established in 1981 that the motion of sound waves in a convergent fluid flow corresponds to the model of the behavior of a quantum field in a classical gravitational field \cite{Unruh1981}. Since those seminal works, Hawking radiation has been considered a fundamental phenomenon of quantum field theory in curved space and has been used as an acid test for quantum theories of gravity \cite{Helfer2003}.

Furthermore, in 1999, S. Corley and T. Jacobson proved for sonic horizons (such as those achieved in BECs) that if the analogues of a black hole (BH) and white hole (WH) horizons are close to each other, the Hawking process is self-amplifying under certain conditions. As the mechanism of amplification resembles that of a laser, this phenomenon was called black hole laser (BHL) \cite{Corley1999a}. The conditions for amplification are a bosonic field, a defined order of the horizons that we call WH--BH order, an anomalous dispersion, and a change of the velocity profile of the fluid flow. 

In particular, models in water, BECs and optics are among the most successful in studying analogue Hawking radiation. In these areas, the theory from which the corresponding dispersion relations is derived is well-known. Also, the study of objects such as white holes is justified, even when there are no known mechanisms of formation in the gravitational case \cite{Faccio2012b}. A numerical study of the hydrodynamical BHL has been recently published \cite{Peloquin2016} In addition, experiments to verify the theoretical models are feasible. For example, in 2014 J. Steinhauer claimed to have measured radiation from a black hole laser formed in a BEC \cite{Steinhauer2014}.

On the other hand, optical analogues are also very successful in proving the quantum properties of Hawking radiation. An optical black hole laser (OBHL) was presented by D. Faccio \textit{et al.} in 2012 and numerical simulations were shown as evidence of the phenomenon \cite{Faccio2012c}. However, the conditions in the optical case are different than those in the sonic one, and therefore, the proof of amplification of the latter cannot be used in the former.

In this work, we present a theoretical description of the OBHL following the approach of Corley and Jacobson, that is, by providing a WKB description of the evolution of frequency modes through a cavity and allowing mode conversion processes at the horizons. In the optical context, the cavity is formed by a pair of light pulses. In particular, we study the Hawking process also for a bosonic field, in this case photons, but in the normal dispersion regime. This forces an inverse order of the horizons (a BH--WH order) to get the proper kinematic behavior. Moreover, the change of velocity is now due to dispersion, and not as a consequence of modifications in the fluid flow. Hence, is Hawking radiation amplified in the optical analogue? And if so, under what conditions? Here we answer these questions.

In addition, we derive the forward propagation of modes and, in this way, the heuristic argument for the amplification is easier to follow. Furthermore, it is known that the amplification depends mainly on the phase difference of the modes in both horizons \cite{Faccio2012c,Leonhardt2007a}. Here, we develop a method to approximate this phase difference and study its behavior. Finally, we present some numerical simulations of the OBHL based on the nonlinear Schr\"odinger equation (NLSE) including negative frequencies \cite{Rubino2012prl,Conforti2013}, which are usually not considered in this kind of simulations but that are necessary to obtain the correct modes of the black hole laser and its amplification \cite{Rubino2012}. Besides, they have been measured in the laboratory \cite{Biancalana2012}.

This work is organized as follows. The concept of the analogue of the event horizon in an optical context is reviewed in Section \ref{analog}. In Section \ref{obhl} we describe the fundamental ideas behind the construction of an OBHL and provide the necessary theory to define the values of the frequency modes that interact in the cavity following the WKB description. Then, the theoretical analysis of the propagation of frequency modes through the cavity and the proof of the amplification of the Hawking process in the OBHL are both detailed in Section \ref{amplification}. Numerical simulations of the propagation of modes in the cavity are shown in Section \ref{num}. Finally, we present our conclusions in Section \ref{conclusions}.

\section{The optical analogue of the event horizon} \label{analog}
All the analogues of the event horizon consider the black hole spacetime as a moving medium, i.e., as a fluid whose movement is caused by gravity, and consider light as waves moving in this fluid. For the optical case the analogy goes one step further: the waves are light waves and the moving fluid is replaced by propagation inside a dielectric material \cite{Philbin2008}. We will now summarize and compare both analogies.

\subsection{Space time as a moving fluid}\label{stasfluid}
\begin{figure}
	\centering
	\includegraphics[width=60mm]{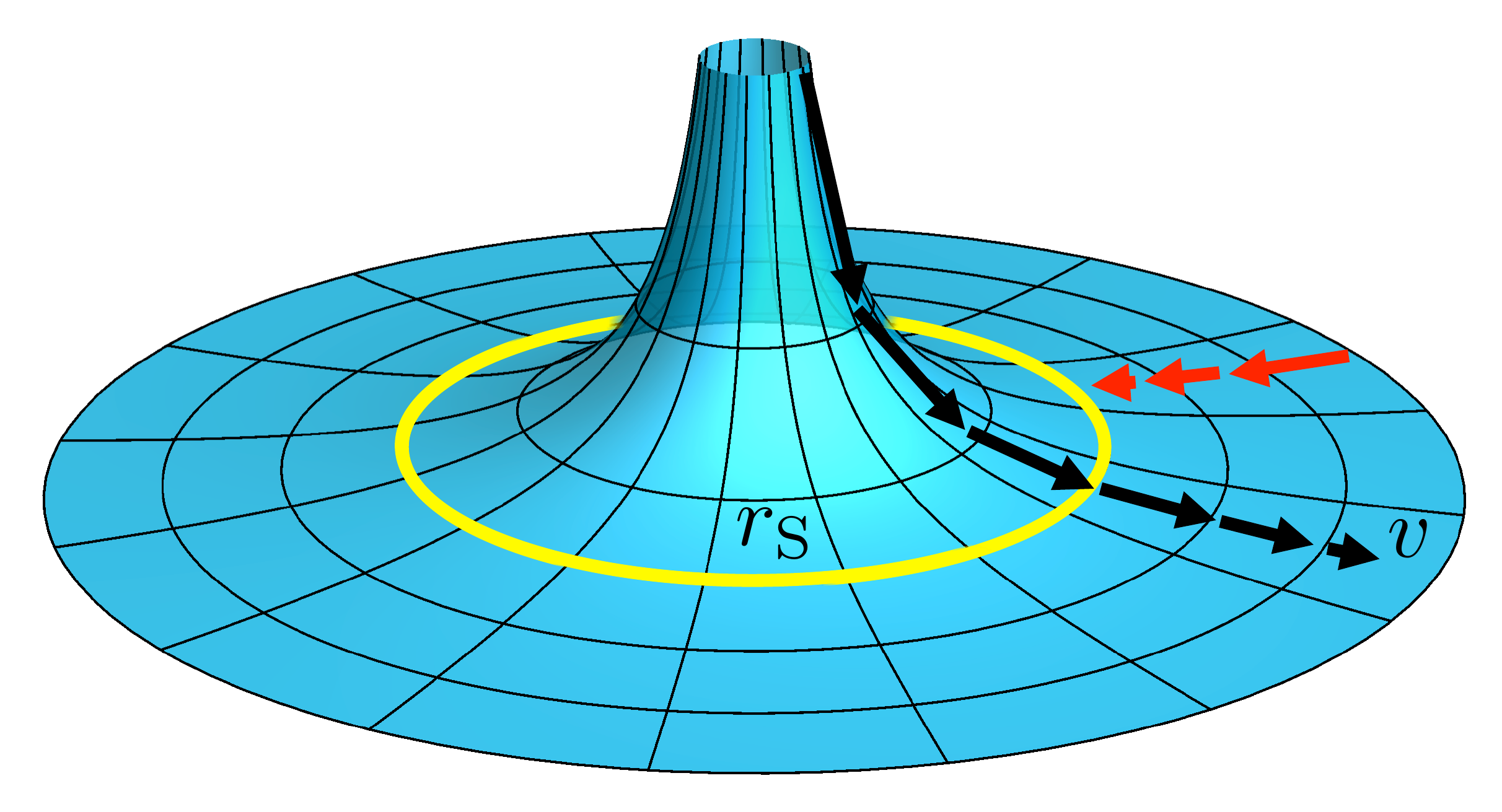}\hspace{15mm}
	\includegraphics[width=60mm]{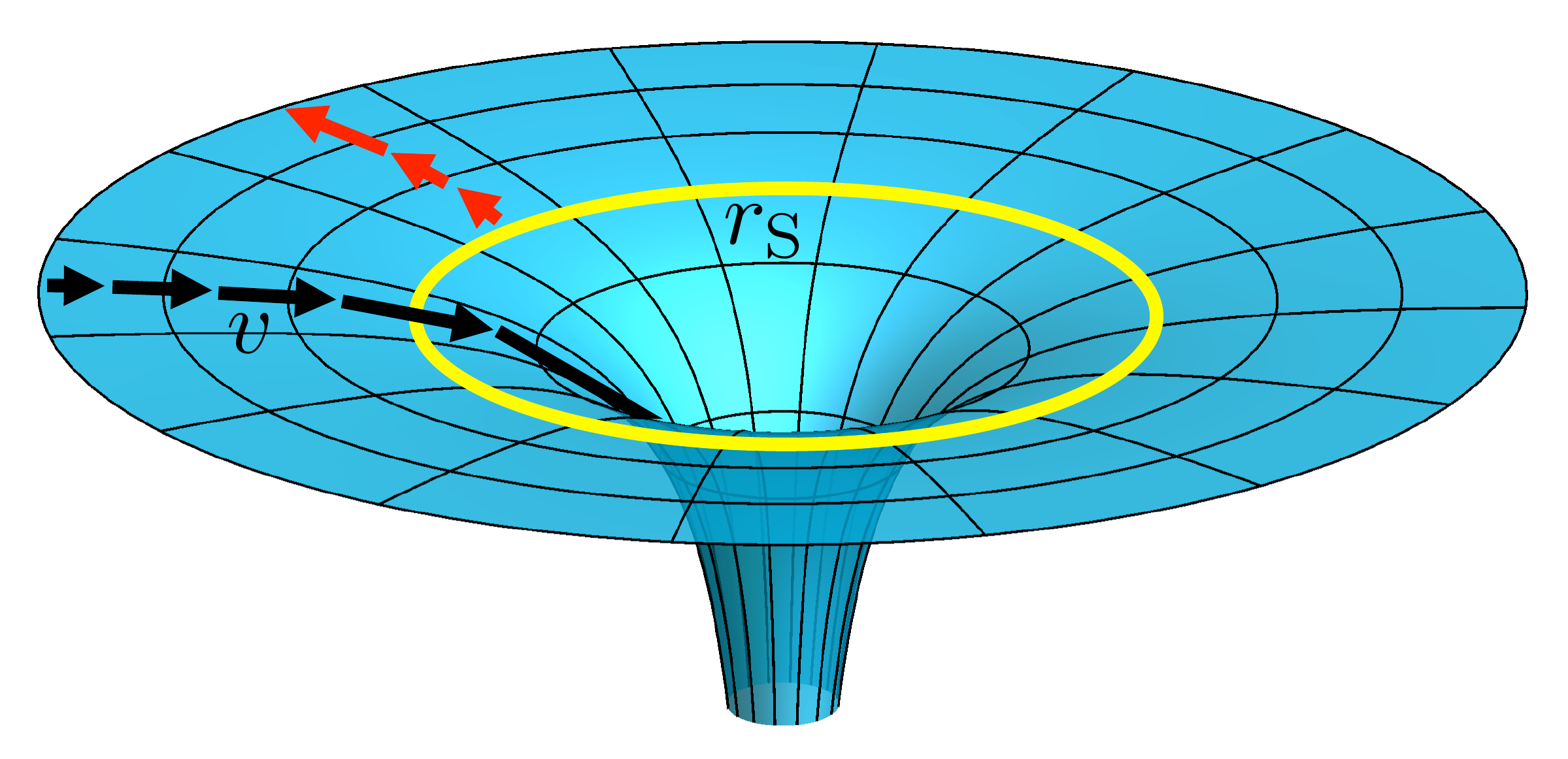}
	\caption{(Color online). Flowing river of space in an astrophysical white hole (left) and black hole (right). At the horizons (yellow lines), the space flow (black arrows) equals $c$. Counter-propagating modes (red arrows) cannot escape from the region inside the event horizon in the black hole or penetrate to the same region in the white hole.}
	\label{geodesics}
\end{figure}

As we are interested in the most basic features of black holes, we choose to study the simplest of them: the one that only has mass $M$ (no charge $Q$ nor angular momentum $L$). This black hole is described by the Schwarzschild metric, which characterizes a spherically-symmetric space with a mass $M$ at the origin. The corresponding metric is given in Painlev\'e-Gullstrand-Lema\^itre coordinates \cite{Painleve1921,Gullstrand1922,Lemaitre1933} by:
\begin{equation}
	\dd s^2= c^2 \dd t^2-\left( \dd r + \sqrt{\frac{r_\text{S}}{r}}c\, \dd t\right)^{2} - r^2\dd\Omega^2,
	\label{schw}
\end{equation}
where $\dd \Omega^2 = \dd \theta^2+\sin^2\theta\, \dd\phi^2$ is the solid-angle element, $(r,\theta,\phi)$ are the spherical coordinates, $c$ is the speed of light in vacuum, and $r_\text{S}=2GM/c^2$ is the Schwarzschild radius. For the Hawking effect and the fluid analogue it is only necessary to consider a $1+1$ dimensional metric, which can be done by setting $\dd\Omega=0$. The term $v(r)=c\sqrt{r_\text{S}/r}$ of Eq. \eqref{schw} can be interpreted as the velocity profile of space if it is thought of as a fluid \cite{Leonhardt2010}. In this picture, space emerges out of the white hole or falls into the black hole as shown in Fig. \ref{geodesics}. At the Schwarzschild radius $r=r_\text{S}$, the flow velocity equals $c$ so the speed of light is exceeded by the flow  beyond that point. Therefore, it is impossible to escape out of the black hole once inside the Schwarzschild radius. In the time reversal case, corresponding to the white hole, it is impossible to penetrate beyond $r_\text{S}$.

\subsection{Optical analogy of a horizon}\label{optana}
There were attempts to create a flow in an optical context in order to induce a kinematic horizon using slow light \cite{Leonhardt2002}. It was later proposed that a localized pulse that propagates through an optical fiber and a change of reference frame are sufficient to recreate the horizon \cite{Philbin2008}.

In terms of the laboratory coordinates $(z,t)$, a retarded time $\tau$ and a propagation time $\zeta$ can be defined as
\begin{equation}
	\tau=t-\frac{z}{u}, \qquad \zeta=\frac{z}{u},
\end{equation}
where $u$ is the group velocity of the localized pulse, which acts as a perturbation. The coordinates $(\zeta,\tau)$ define the co-moving frame. We will consider a fixed-shape pulse propagating in $z$ with constant velocity. Then, the properties of the effective medium in the co-moving frame depend only on $\tau$, which plays the role of space (whereas $\zeta$ plays the role of time). This can be achieved exactly by using solitons or as an approximation if the horizon dynamics is faster than the pulse dynamics in the fiber.

One of the main advantages of working in the co-moving frame is that the frequency in this frame $\omega'$ is invariant \cite{Bermudez2016a}. Actually, it can be shown that frequency conservation in the co-moving frame is equivalent to momentum conservation in the laboratory frame, and in turn, to a phase-matching condition \cite{Agrawal2013}. The frequency $\omega'$ is explicitly given by the Doppler relation
\begin{equation}
	\omega'(\omega)=\omega - \omega \frac{u}{c}n(\omega).
	\label{wprimaw}
\end{equation}
The effect of the localized pulse is to change the original phase index of the fiber $n(\omega)$ by an additional contribution $\delta n$, which depends on the intensity of the pulse, i.e., $\delta n \propto I(\tau)$, and moves inside the fiber with a constant velocity $u$. This is called the optical Kerr effect \cite{Philbin2008,Agrawal2013}. The effective phase index $n_\text{eff}$ is then given by
\begin{equation}
	n_\text{eff}(\omega,\tau)=n(\omega)+\delta n(\tau).
	\label{refindex}
\end{equation}

\subsection{Interaction of a probe pulse with a light perturbation}
\begin{figure}
	\centering
	\includegraphics[width=70mm]{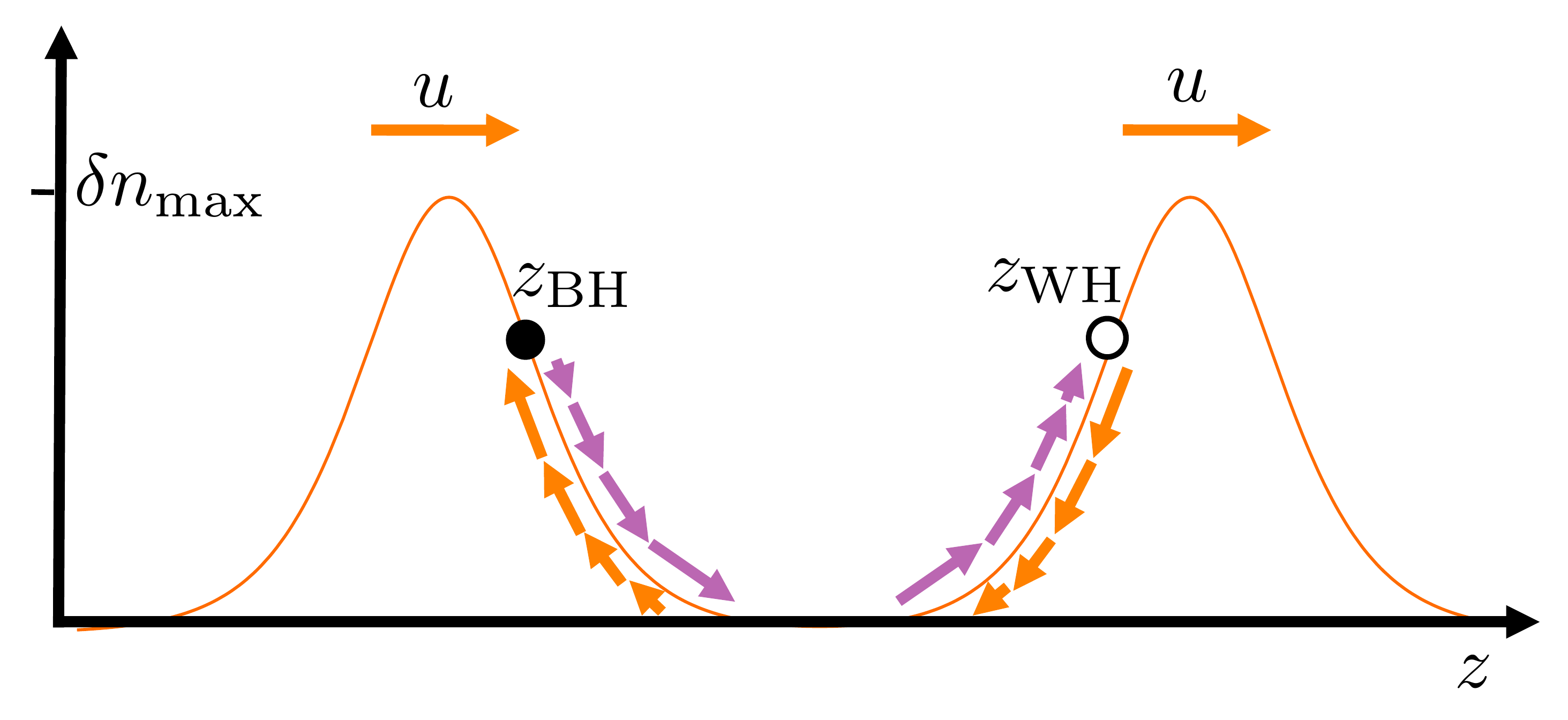}\hspace{10mm}
	\includegraphics[width=70mm]{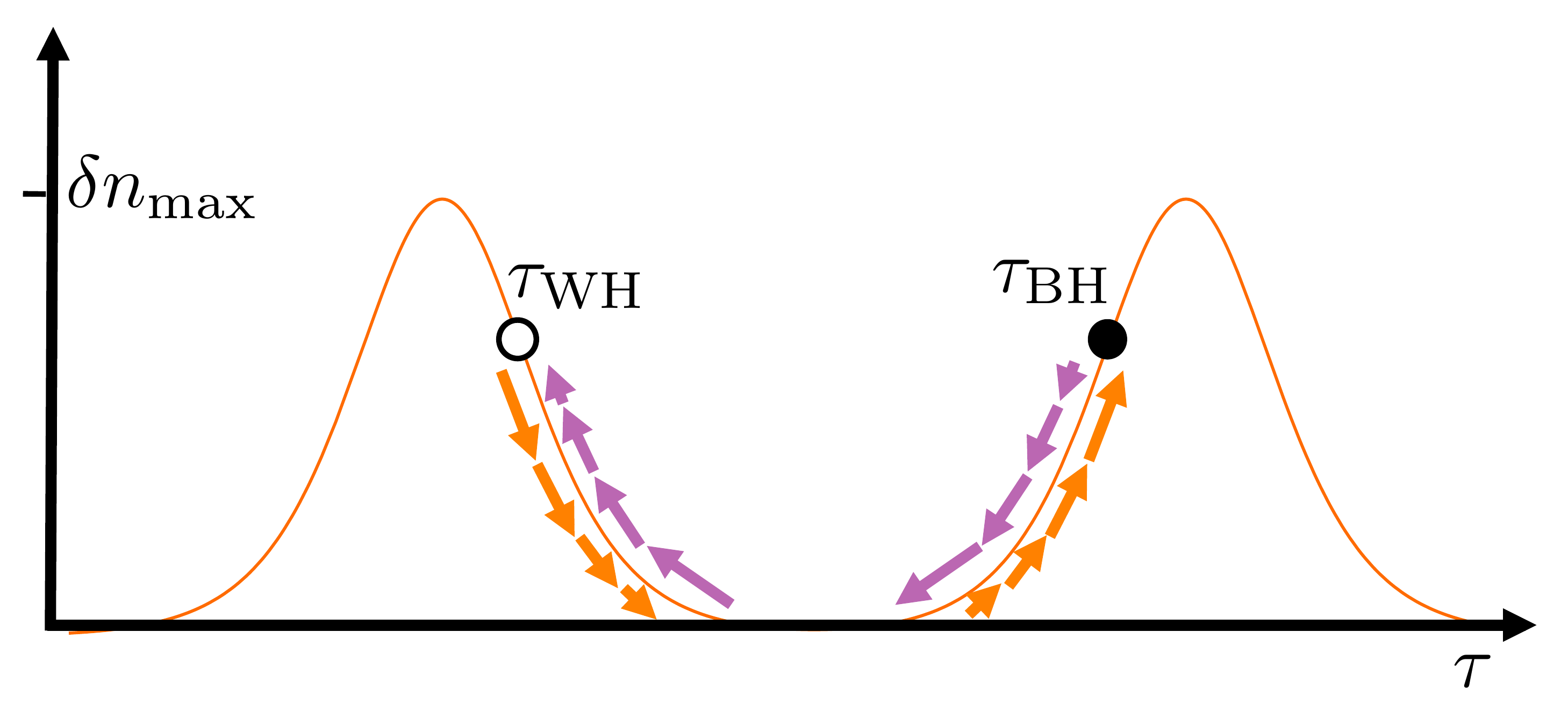}
	\caption{(Color online). (Left) Cavity formed by two solitons in the laboratory frame. The white hole (WH) and black hole (BH) horizons are in the trailing and leading edge of the pulses, respectively. (Right) Location of the horizons in the co-moving frame, with an apparent flip of the horizons. The probe pulse is blocked at the WH and is unable to escape from the BH (magenta arrows) due to the flow velocity (orange arrows indicate the magnitude in the co-moving frame).}
	\label{cavity}
\end{figure}

Given a localized perturbation moving through an optical fiber with velocity $u$, we consider a probe pulse with velocity $v$ in the fiber. Two situations may occur according to the values of $u$ and $v$. In the case where $u<v$, the perturbation will be caught up by the probe pulse approaching from behind, as viewed in the laboratory frame. As the probe pulse reaches the trailing edge of the perturbation, it will be slowed down by the higher group index, and it will eventually be blocked at a point $z_\text{WH}$ in the laboratory frame or $\tau_\text{WH}$ in the co-moving one. The following relation is then satisfied
\begin{equation}
	u=\frac{c}{n_{g}+\delta n(\tau_\text{WH})},
	\label{whhorizonn}
\end{equation}
where $n_g$ is the group index of the fiber, which is inversely proportional to the group velocity $v_g$. This is reminiscent of a white hole horizon. On the other hand, if $u>v$, an analogue of a black hole horizon is obtained for the leading edge of the perturbation. Therefore, under these conditions, a single perturbation recreates the analogue of both a white hole (trailing edge) and a black hole (leading edge). In addition, it should be noted that there is an apparent flip between left and right when we transform between the laboratory and the co-moving frames, as shown in Fig. \ref{cavity}.

The condition of occurrence of the event horizons is given by the relation 
\begin{equation}
	\frac{c}{n_{g}+\delta n_\text{max}}\leq u \leq \frac{c}{n_{g}},
	\label{horcond}
\end{equation}
where $\delta n_\text{max}$ is the maximum change in the phase index, which is reached at the peak of the perturbation. There are two possible ways of interpreting Eq. \eqref{horcond}. If the background group index $n_g$ and perturbation amplitude $\delta n_\text{max}$ are given, only perturbations that move at a velocity $u$ in the interval defined by Eq. \eqref{horcond} give rise to both analogue horizons. On the other hand, if $u$ and $\delta n_\text{max}$ are given, the effect of the horizons will occur for pulses with frequencies $\omega$ such that $n_g(\omega)$ satisfies Eq. \eqref{horcond} \cite{Faccio2012b}.

If we consider the effect of the perturbation, $\omega'$ is modified by the effective phase index as
\begin{equation}
	\omega'_\text{eff}(\omega)=\omega - \omega \frac{u}{c}\left[n(\omega)+\delta n\right]
	=\omega'(\omega)-\omega\frac{u}{c}\delta n, \label{Cofreq}
\end{equation}
where $\omega'(\omega)$ is given by Eq. \eqref{wprimaw}. Notably,  $\omega'_\text{eff}$ is the conserved quantity in the system formed by the probe and the perturbation. We will use this fact in the derivation of the next section.

\section{The optical black hole laser}\label{obhl}
The basic idea of an optical black hole laser is to trap a probe pulse in a cavity formed by light perturbations. As previously mentioned, this probe will be confined because it will be slowed down when it reaches any of the perturbations due to the Kerr effect. A cavity can be constructed by using two positive-valued perturbations generated by two independent laser pulses that are placed a certain distance $z$ (or time $\tau$) apart inside the fiber \cite{Faccio2012c}, as shown in Fig. \ref{cavity}.

In order to construct a cavity that does not modify its shape during $\zeta$-propagation (so as to isolate the effect of the local decrease in the group velocity), we use two solitons centered at the same frequency $\omega_c$ and separated a fixed time $\tau_c$ ($c$ stands for cavity). Usually, the dispersion of the fiber is in the so-called normal dispersion regime, i.e., $n_g(\omega)$ increases with increasing $\omega$. Nevertheless, solitons can only exist in the anomalous regime, in which $n_g(\omega)$ decreases with increasing $\omega$. Thus, there must be some frequency interval in the dispersion of the fiber that is anomalous. This situation is very common for example in photonic crystal fibers \cite{Agrawal2013}.

\subsection{Determination of frequency modes}
\begin{figure}
	\centering
	\includegraphics[width=85mm]{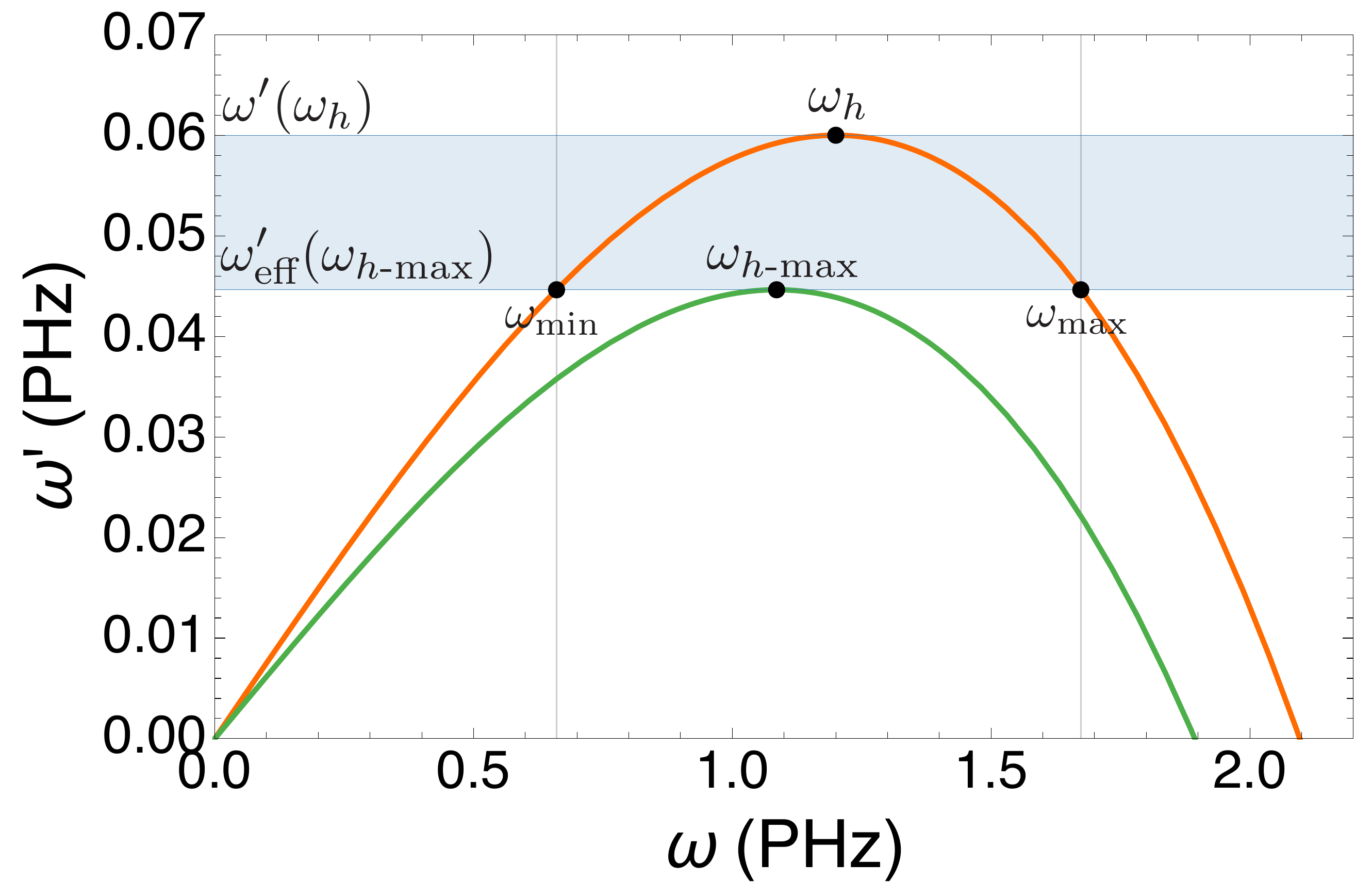}
	\caption{(Color online). The frequencies for which the horizons occur are found by solving Eq. \protect\eqref{hor2} for $\delta n=0$ (orange) and $\delta n=\delta n_{\text{max}}$ (green), yielding the frequencies $\omega_{h}$ and $\omega_{h\text{-max}}$, respectively. The laboratory frequencies in the interval $(\omega_\text{min},\omega_\text{max})$ experience the effect of the horizons as its corresponding value of the co-moving frequency is in the interval $\left(\omega'_{\text{eff}}\left(\omega_{h\text{-max}}\right),\omega'(\omega_{h})\right)$. }
	\label{restrictions}
\end{figure}

The range of frequencies that experience the horizons is found by solving Eq. \eqref{horcond}. Alternatively, the frequencies can be obtained by determining the null values of the derivative of $\omega'_\text{eff}(\omega)$ in Eq. \eqref{Cofreq}, that is,
\begin{equation}
	\frac{\text{d}\omega'_\text{eff}}{\text{d}\omega}=\frac{\text{d}\omega'}{\text{d}\omega}-\frac{u}{c}\delta n=0. \label{hor2}
\end{equation} 
Then, the frequencies that correspond to the analogue of the event horizons depend on the value of $\delta n$. The limiting values of Eq. \eqref{horcond} can be obtained by setting $\delta n=0$ and $\delta n=\delta n_{\text{max}}$. In particular, for regions of normal dispersion with positive values of $\omega'_{\text{eff}}$, the solutions correspond to maxima of $\omega'_\text{eff}(\omega)$, so if $\omega_{h}$ is the solution of Eq. \eqref{hor2} corresponding to $\delta n=0$ and $\omega_{h\text{-max}}$ corresponds to $\delta n=\delta n_{\text{max}}$, then the range of frequencies that experience the horizons is such that the corresponding frequencies in the co-moving frame are contained in the interval $\left(\omega'_\text{eff}(\omega_{h\text{-max}}), \omega'(\omega_{h}) \right)$, as shown in Fig. \ref{restrictions}. The frequency interval in the laboratory frame is thus given by the solutions of $\omega'(\omega)= \omega'_{\text{eff}}(\omega_{h\text{-max}})$. In Fig. \ref{restrictions}, we called $\omega_\text{min}$ and $\omega_\text{max}$ the frequencies that satisfy this condition, so the laboratory frequencies that experience the horizons are contained in the interval $(\omega_\text{min}, \omega_\text{max})$.

The dynamical evolution of pulses in this interval corresponds to a region of normal dispersion. Given this region, a simplified model of the dispersion relation (see Ref. \cite{Bermudez2016a}) can be used to study the effect of the perturbation. This model includes the essential properties to describe the dispersion of light in a fiber, and is able to create the analogue of the event horizons. In the absence of perturbation, the propagation constant $\beta(\omega)$ can be approximated by the relation
\begin{equation}
	\beta^{2}(\omega)=\frac{\omega^{2}}{c^{2}}\left(b_1^2+b_2\omega^2\right), \label{betaapp}
\end{equation} 
where the parameters $b_1$ and $b_2$ are found by imposing that $\omega'(\omega_{h})$ is a maximum of the dispersion relation.

The frequency modes that conserve $\omega'$ are the ones that can exist after a nonlinear interaction with the cavity that causes the mode conversion processes. We can find these modes by solving Eq. \eqref{Cofreq} using the approximated model for $\beta(\omega)$ given by Eq. \eqref{betaapp}. For a given input frequency $\omega_\text{IN}$, the co-moving frequency is $\omega'_0\equiv\omega'(\omega_\text{IN})$, that is
\begin{equation}
	\omega'_0=\omega-u\frac{\omega}{c}\left( \sqrt{b_1^2+b_2\omega^2}+\delta n \right). \label{Freqmodes}
\end{equation}
As we mentioned in Section \ref{optana}, $\omega'_{0}$ is a conserved quantity. On the other hand, this equation shows the two main differences between the black hole laser in the optical and sonic contexts (see Eqs. (2.6) and (2.7) in Ref. \cite{Corley1999a}). First, the dispersion relation for the OBHL is normal, in contrast with the anomalous dispersion considered for the sonic case. In addition, in the sonic case there is a velocity profile $v(x)$ that yields the solutions of Eq. \eqref{Freqmodes} by fixing a value of $v(x)$ that corresponds to a superluminal or a subluminal velocity. In contrast, in the optical case there is no velocity profile because the velocity $u$ of the frame co-moving with the perturbation is fixed. The corresponding modes are found by the effect of the perturbation which is changing the dispersion relation by an additional contribution $\delta n$ to the phase index.

\begin{figure}
	\centering
	\includegraphics[height=50mm]{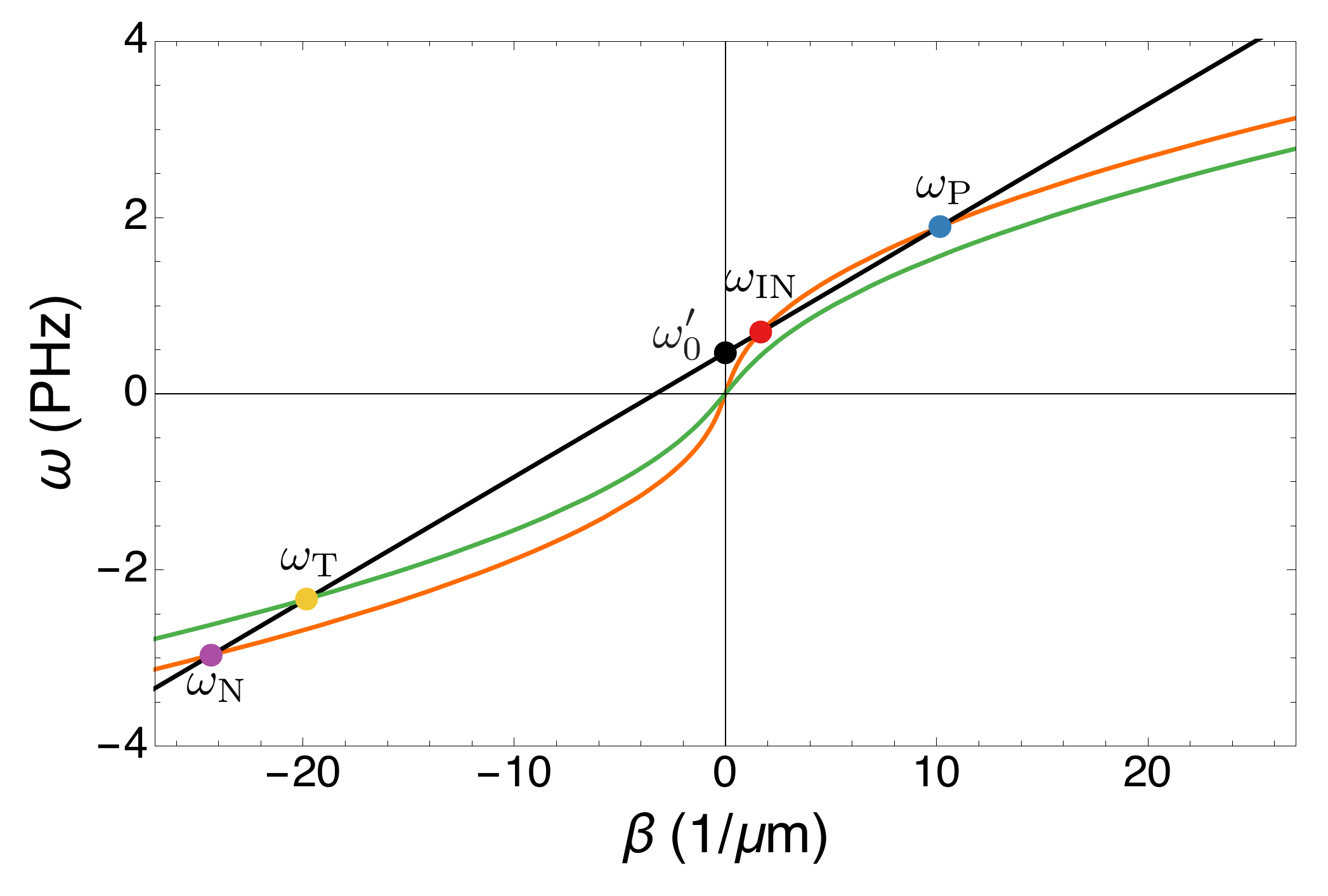}\hspace{5mm}
	\includegraphics[height=50mm]{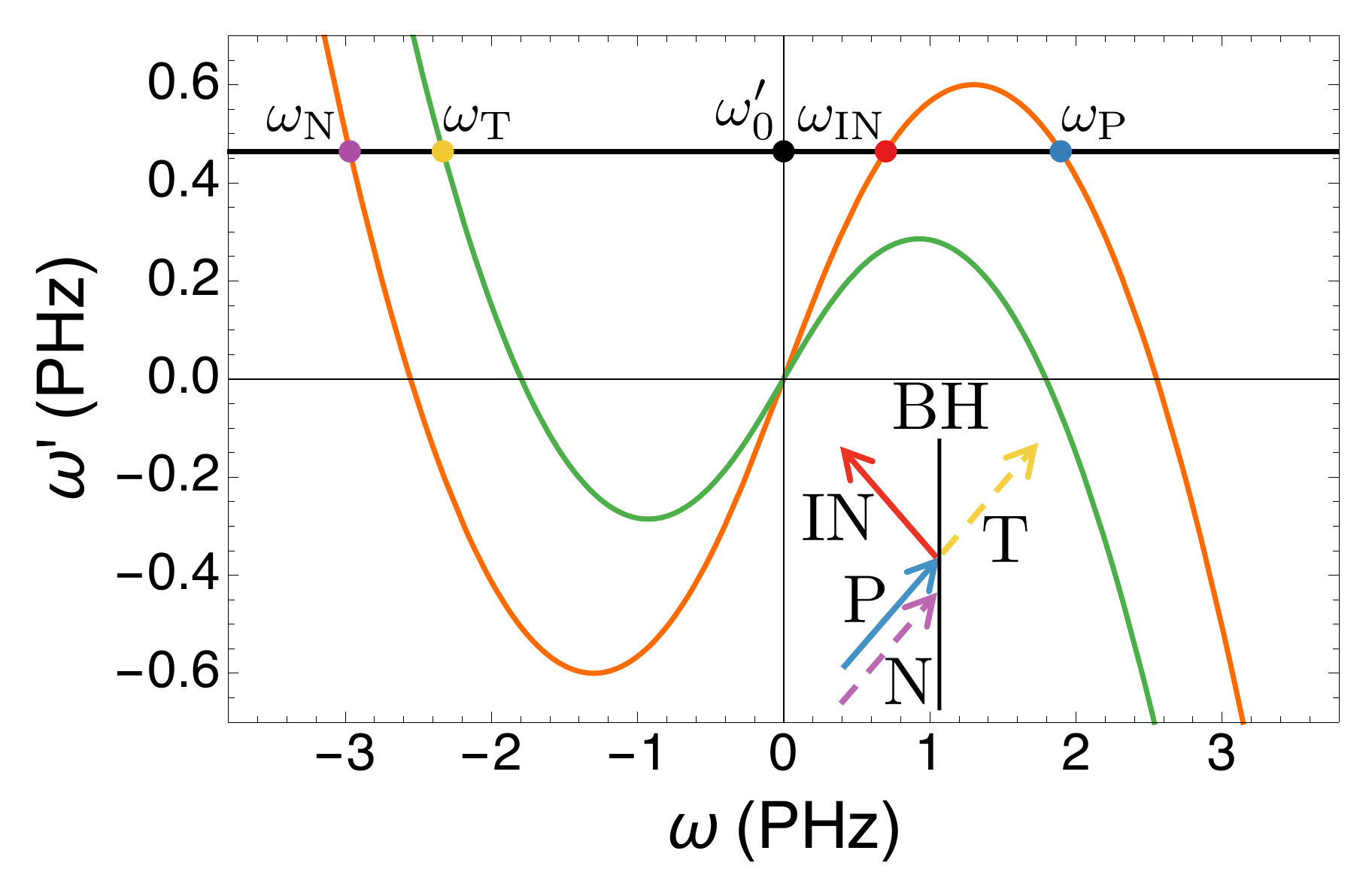}
	\caption{(Color online). Modes corresponding to a fixed value of the co-moving frequency $\omega'_0$ in the laboratory frame (left) and in the co-moving frame (right). The three modes $\text{IN}$, $\text{P}$, and $\text{N}$ are found for the dispersion relation with $\delta n=0$ (orange line). For the dispersion with $\delta n=\delta n_\text{max}$ (green line), only the $\text{T}$ mode is found. The inset shows mode propagation in the vicinity of a BH.}
	\label{ModeSolutions}
\end{figure}

In addition to the input mode (IN) with frequency $\omega_\text{IN}$, two other solutions of Eq. \eqref{Freqmodes} are found for $\delta n=0$. They will be denoted as $\text{P}$ (positive laboratory frequency mode) and $\text{N}$ (negative laboratory frequency mode). These three modes are trapped in the cavity formed by the perturbations and, according to Eq. \eqref{horcond}, correspond to a superluminal velocity. On the other hand, the effect of the contribution $\delta n$ can be thought of as a clockwise rotation of the curve $\omega'(\omega)$. Therefore, for a large enough $\delta n=\delta n_\text{max}$ there is only one additional solution that shall be denoted as $\text{T}$ (transmitted mode) that exists outside the cavity and corresponds to a subluminal velocity. The four possible solutions are shown in Fig. \ref{ModeSolutions}.
The number of solutions for the superluminal and subluminal velocities shows another difference between the optical and sonic cases. If the possible solutions of the frequency modes in the OBHL were found by fixing the values of a velocity profile, as in the sonic case, then the number of solutions in the subluminal and superluminal would be inverted.

The approach of Corley and Jacobson \cite{Corley1999a} can be followed by noticing that a rearrangement of Eq. \eqref{Freqmodes} yields
\begin{equation}
	\left[ -\mathcal{W}_0+\mathcal{W}v(\tau)\right]^{2}=\mathcal{W}^{2}+\frac{\mathcal{W}^{4}}{\Omega_0^{2}}, \label{rearr}
\end{equation}
where $\mathcal{W}=b_1 \omega$, the constants $\mathcal{W}_0$ and $\Omega_0$ are given by the relations $\mathcal{W}_0=\omega'_0c/u$, $\Omega_0^{2}=b_1^4/b_2$ and we define the quantity
\begin{equation}
	v(\tau)=\frac{c}{b_1 u}\left[1-\frac{u}{c}\delta n(\tau)\right].
	\label{eqvtau}
\end{equation}
The importance of the rearrangement shown in Eq. \eqref{rearr} is that its structure is of the same form as that used in the sonic case. So, even if the frequency modes are not given by the possible values of a velocity profile, they can be found according to the values of the quantity $v(\tau)$, thus establishing an equivalence between the velocity profile in the sonic case and the modification of the dispersion relation by the additional contribution $\delta n$ in the optical case.

\subsection{Propagation of frequency modes}\label{propa}
The group velocity of each mode can be found using Eq. \eqref{hor2}. The direction can easily be obtained with the sign of the derivative in the $\omega'(\omega)$ diagram (see Fig. \ref{ModeSolutions}). The modes that correspond to $\delta n=0$ propagate through the interior of the cavity in such way that the IN mode travels in the BH$\rightarrow$WH direction, whereas P and N propagate in the opposite direction. On the other hand, the T mode exists only outside the cavity, when $\delta n=\delta n_\text{max}$. This mode propagates away from the BH to the right or approaches the WH from the left, as shown in the inset of Fig. \ref{ModeSolutions}. Strictly speaking, after this mode moves away from the localized perturbation it transforms into an N mode (which also has negative-frequency and that moves in the same direction) and leaves the system. In our study we will continue labeling this mode as T to distinguish it from the N mode inside the cavity.

\begin{figure}
	\centering
	\includegraphics[height=45mm]{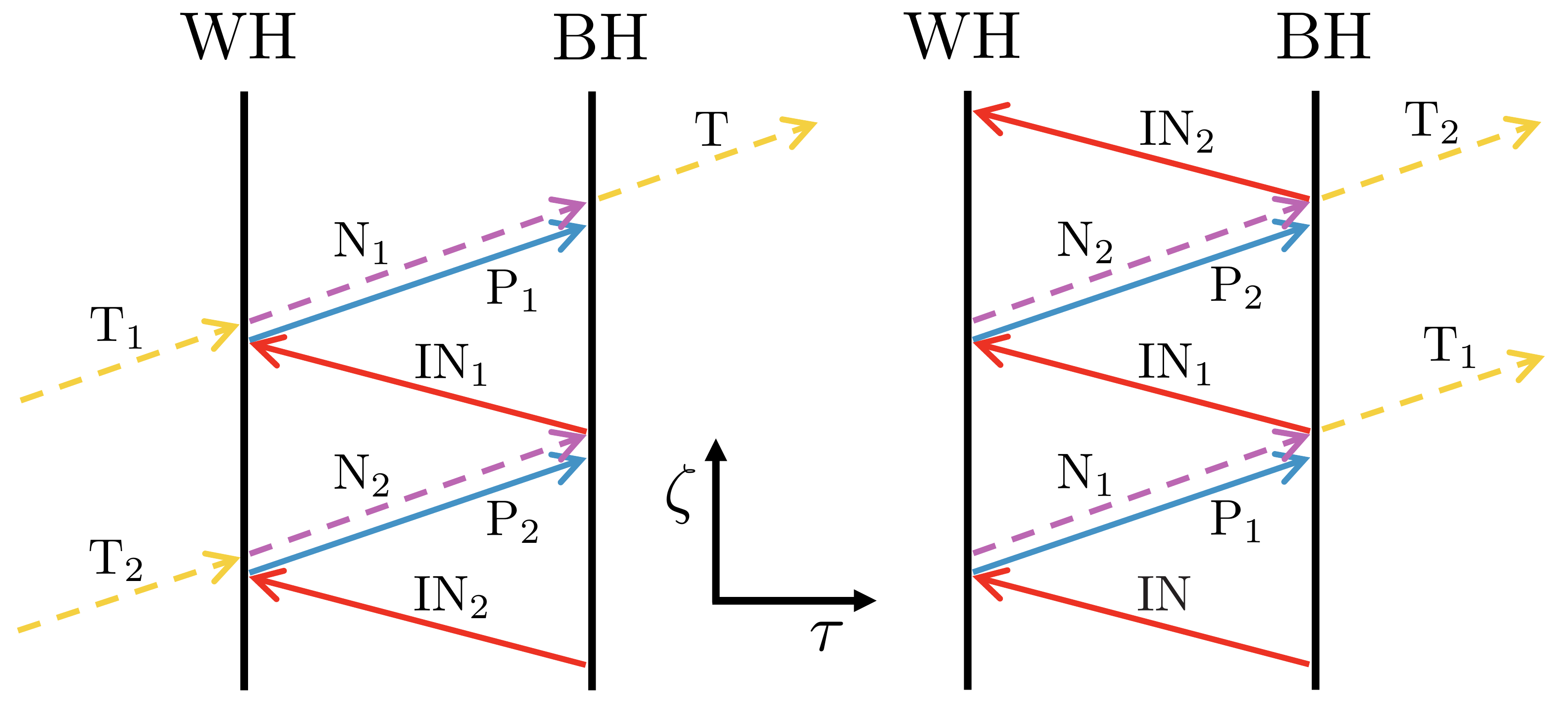}
	\caption{(Color online). Spacetime evolution of an outgoing T mode backwards in $\zeta$ (left) and of an incoming IN mode forwards in $\zeta$ (right). Destruction of wave modes by mode conversion is indicated by the end of lines. Solid lines refer to positive frequency (and norm) modes and dashed lines to negative ones.}
	\label{evolution}
\end{figure}

In particular, the evolution of a final T mode backwards in $\zeta$ can be described based on a WKB framework complemented by mode conversion processes at the horizons. In this picture, as shown in the diagram of Fig. \ref{evolution} (left), the final T mode approaches the BH, where it transforms into an N mode. In addition, mode conversion allows the existence of a $\text{P}$ mode that is not obvious based only on the dispersion curve, but that is possible because this mode conserves $\omega'_0$ (see Fig. \ref{ModeSolutions}). Even though the $\text{IN}$ mode also conserves the frequency $\omega'_0$, there is no mode conversion to this mode because it moves in the opposite direction. Then, the pair of P and N modes propagate towards the WH. When they reach it, the two modes give origin to a T mode that propagates further to the left of the cavity, and an IN mode that propagates again to the BH. As the IN mode approaches the BH, a pair of P and N modes arises through mode conversion and the subsequent processes are as previously described.

On the other hand, let us consider the evolution of an IN mode inside the cavity forwards in $\zeta$, as seen in Fig. \ref{evolution} (right). This mode travels towards the WH, where it is converted into a pair of $\text{P}$ and $\text{N}$ modes by mode conversion. Conversion to the N mode with negative frequency is possible because this mode also conserves $\omega'_{0}$ (see Fig. \ref{ModeSolutions}), and moves in the proper direction unlike the T mode. Then, these modes travel towards the BH, where they give origin to a $\text{T}$ and an $\text{IN}$ modes. At this stage, the T mode escapes the cavity and the $\text{IN}$ mode repeats the process.

\subsection{Approximation of frequency modes}
We start by proposing the following action for the bosonic field in the $(\tau, \zeta)$ coordinates
\begin{equation}
	S=\int{\dd \zeta \mathcal{L}}=\frac{\epsilon_{0}}{2}\int{\dd\zeta\dd\tau\left\lbrace \left[ \left( \partial_{\zeta}-v(\tau)\partial_{\tau}\right)\Phi\right]^{2}+\Phi\hat{F}(\partial_{\tau})\Phi\right\rbrace}, \label{action}
\end{equation}
where the operator $\hat{F}(\partial_{\tau})$ is defined as 
\begin{equation}
	\hat{F}=\partial_{\tau}^{2}-\frac{\partial_{\tau}^{4}}{\Omega_0^{2}}
\end{equation}
Given this action, the Euler-Lagrange equations yield the following equation of motion
\begin{equation}
	\left[\partial_{\zeta}-\partial_{\tau}v(\tau)\right]\left[\partial_{\zeta}-v(\tau)\partial_{\tau}\right]\Phi=\hat{F}(\partial_{\tau})\Phi.\label{motion}
\end{equation}
Proposing a solution of the form 
\begin{equation}
	\Phi(\tau,\zeta)=\exp\left(i\mathcal{W}_0\zeta+i\int{\dd \tau \, \mathcal{W}(\tau)} \right), \label{solution}
\end{equation}
and neglecting derivatives of both $\mathcal{W}(\tau)$ and $v(\tau)$, we reproduce the dispersion relation of Eq. \eqref{rearr}. On the other hand, allowing a more general solution of the form $\Phi(\tau,\zeta)=\exp\left(i \mathcal{W}_0\zeta \right)\phi(\tau)$, substitution in Eq. \eqref{motion} yields the following relation
\begin{equation}
	-\frac{1}{\Omega_0^2}\phi^{(iv)}(\tau)+\left[1-v^{2}(\tau)\right]\phi''(\tau)+2v(\tau)\left[i\mathcal{W}_0-v'(\tau)\right]\phi'(\tau)-i\mathcal{W}_0\left[i\mathcal{W}_0-v'(\tau)\right]\phi(\tau)=0. \label{motion2}
\end{equation}
Following the WKB method, we choose $\phi(\tau)$ as the $\tau$-term in Eq. \eqref{solution} and obtain the following expression for $\mathcal{W}(\tau)$
\begin{align}
	\begin{split}
		-\frac{1}{\Omega_0^{2}}\mathcal{W}^{4}&-\left[1-v^{2}(\tau)\right]\mathcal{W}^{2}-2v(\tau)\mathcal{W}_{0}\mathcal{W}+\mathcal{W}_{0}^{2}\\
		&=-i\frac{\dd}{\dd \tau}\left\{\frac{2}{\Omega_0^{2}}\mathcal{W}^{3}+\left[1-v^{2}(\tau)\right] \mathcal{W}^{2}+\mathcal{W}_{0}v(\tau)\right\} -\frac{1}{\Omega_0^2} \left[4\mathcal{W}\mathcal{W}''+3\left(\mathcal{W}''\right)^{2}\right]+\frac{i}{\Omega_{0}^{2}}\mathcal{W}^{(3)}.
	\end{split}\label{wkb}
\end{align}

After letting $\tau \rightarrow \alpha \tau$ and assuming that $\mathcal{W}(\tau)$ may be expanded in inverse powers of $\alpha$, (where $\alpha$ is just an auxiliary parameter that will be set as $\alpha=1$ at the end), the condition that each power of $1/\alpha$ vanishes separately gives an infinite set of equations, where the lowest orders produce the following frequencies of the modes in consideration
\begin{align}
	\mathcal{W}_{\text{IN},\text{T}}=&-\frac{\mathcal{W}_{0}}{1-v(\tau)}
	,\label{INT}\\
	\mathcal{W}_{\text{P,N}} =& \pm \Omega_{0}\sqrt{v^{2}(\tau)-1}+\frac{\mathcal{W}_{0}v(\tau)}{1-v^{2}(\tau)}+i\frac{3}{4}\frac{\dd}{\dd \tau}\ln\left[v^{2}(\tau)-1 \right]
	.\label{PN}
\end{align}
These solutions should be evaluated with $\delta n=0$ for the IN, P, and N modes and with $\delta n=\delta n_\text{max}$ for the T mode.

\section{Hawking Amplification in the OBHL}\label{amplification}
In this section we will obtain the connection formulas, which describe the relation between the modes at both sides of the horizons. In addition, we will derive the equations for the evolution of packets in the OBHL and we will prove the amplification of the Hawking radiation. Finally, we will use a simple model to study the conditions of maximal amplification.

\subsection{Connection formulas}
\begin{figure}
	\centering
	\includegraphics[width=80mm]{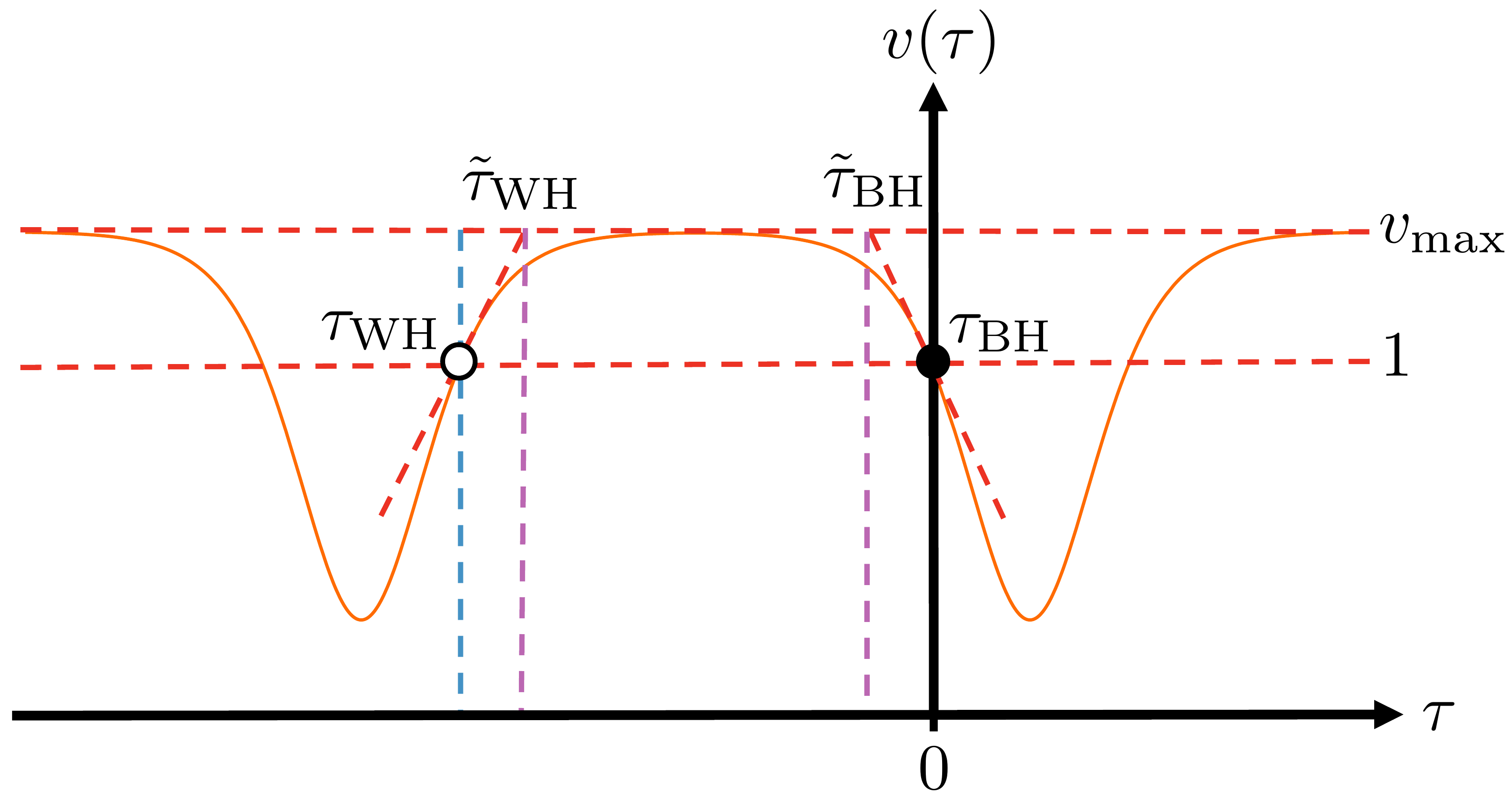}	
	\caption{(Color online). Equivalent velocity profile $v(\tau)$ in the case of a cavity formed by two solitons. The points $\tau_{\text{WH}}$ and $\tau_{\text{WH}}$ are defined as the solutions of $v(\tau)=1$. Note that $\tilde{\tau}_\text{BH}$ and $\tilde{\tau}_\text{WH}$ are important times in the approximation the phase difference in Section \protect\ref{secphase}.}
	\label{vtau}
\end{figure}

We can set the origin of our $\tau$-coordinate so that $\tau_\text{BH}=0$, and work with $b_{1}=1$ in a set of units such that $v(0)=1$ in Eq. \eqref{eqvtau}. This arrangement can be seen in Fig. \ref{vtau}. Under this condition, we expand $v(\tau)$ up to first-order, which is useful to define a parameter $\kappa$ that plays the role of the surface gravity as it is common in analogue systems \cite{Philbin2008,Corley1999a}. The parameter $\kappa$ is explicitly given by
\begin{equation}
	\kappa=\left.\frac{\text{d}v}{\text{d}\tau}\right|_{\tau=0}=-\frac{1}{b_1}\left.\frac{\text{d} \delta n(\tau)}{\text{d}\tau} \right|_{\tau=0}, \label{kappap}
\end{equation}
and is a negative quantity for the leading edge where the BH is located. Following the approach of Corley \cite{Corley1998}, the Laplace transform method can be used to solve Eq. \eqref{motion2}. Under the assumptions that $|\kappa/\tau|\ll 1$ and $|\kappa_1^2/\kappa|\ll 1$ (where $\kappa_{1}$ is the second term in the expansion of $v(\tau)$ around $\tau=0$), Eq. \eqref{motion2} takes the form 
\begin{equation}
	-\Omega_0^{-2}\phi^{(iv)}(\tau)-2\kappa\tau\phi''(\tau)+2\left(i \mathcal{W}_{0}-\kappa\right)\phi'(\tau)-i\mathcal{W}_{0}\left(i\mathcal{W}_{0}-\kappa\right)=0
\end{equation}
The solutions derived from the Laplace method can be expressed in terms of those obtained by the WKB approach (see Eqs. (\ref{wkb}-\ref{PN})). From there, the following connection formulas hold
\begin{align}
	K\left(\text{e}^{-\pi \mathcal{W}_{0}/(2|\kappa|)}\phi_{\text{P}}+\text{e}^{\pi \mathcal{W}_{0}/(2|\kappa|)}\phi_{\text{N}}\right) &\longleftrightarrow\phi_{\text{T}},\label{connect1}\\
	-\phi_{\text{IN}}+K\left(\text{e}^{\pi \mathcal{W}_{0}/(2|\kappa|)}\phi_{\text{P}}+\text{e}^{-\pi \mathcal{W}_{0}/(2|\kappa|)}\phi_{\text{N}} \right) &\longleftrightarrow 0, \label{connect2}
\end{align}
where the parameter $K$ is defined as 
\begin{equation}
	K=\sqrt{\frac{\mathcal{W}_0}{2\Omega_0}\text{sinh}^{-1}\left(\frac{\pi \mathcal{W}_{0}}{|\kappa|}\right)},
\end{equation}
and the arrows indicate a connection between the solutions at both sides of the horizon. It can be shown \cite{Corley1999a} that the solutions of Eq. \eqref{motion} around the WH, located at $\tau_\text{WH}$, are the complex conjugates of those at the BH for a symmetric cavity, so the corresponding connection formulas for the WH have the same structure as those shown in Eqs. \eqref{connect1} and \eqref{connect2}. More importantly, the solutions at both horizons differ in a phase only, so they can be expressed without loss of generality as 
\begin{align}
	\tilde{\phi}_{\text{IN}}(\tau_\text{WH}-\tau) &= \text{e}^{i\theta_{\text{IN}}(\mathcal{W}_0)}\phi_{\text{IN}}(\tau), \\
	\tilde{\phi}_{\text{P,N}}(\tau_\text{WH}-\tau)&= \text{e}^{i\theta_{\text{P,N}}(\mathcal{W}_0)}\phi_{\text{P,N}}(\tau). 
\end{align}

\subsection{Norm of the frequency modes}
As the generalized Lagrangian density $\mathcal{L}$ of the action \eqref{action} is invariant with respect to the transformation $\phi'=\text{e}^{i\lambda}\phi$ of the complex field $\phi$, an associated current is conserved such that its $\zeta$-component can be used to define the following inner product
\begin{equation}
	\left(\phi_{1},\phi_{2}\right)=\frac{i\epsilon_{0}c^{2}}{u\hbar}\int{\dd \tau \left[ \phi_{1}^{*}\left(\partial_{\zeta}-v(\tau)\partial_{\tau}\right)\phi_{2}-\phi_{2}\left(\partial_{\zeta}-v(\tau)\partial_{\tau}\right)\phi_{1}^{*}\right]},\label{Norm}
\end{equation}
which satisfies $\partial_{\zeta}(\phi,\phi)=0$, that is, the norm is conserved. This inner product coincides with that defined in Ref. \cite{Philbin2008} except for a factor of $-1$. This sign is usually chosen so that the signs of the norm and the laboratory frequency coincide. Norm conservation will prove essential to understand the amplification of the analogue of Hawking radiation in the OBHL. We can construct wave packets from the solutions of the frequency modes as
\begin{equation}
	\psi=\int{\frac{\dd \mathcal{W}_{0}}{\sqrt{\mathcal{W}_{0}}}G_{\mathcal{W}_{0}}\text{e}^{i\mathcal{W}_{0}\zeta}\phi\left[\mathcal{W}(\mathcal{W}_{0}) \right]},
\end{equation}
where $G_{\mathcal{W}_{0}}$ is the amplitude obtained from matching the connection formulas in Eqs. \eqref{connect1} and \eqref{connect2} and the evolution formulas. The latter describe the behavior of wave packets as they evolve with respect to $\zeta$ and will be obtained in the following section. We will also show that the sign of the norm of the wave packet agrees with the sign of the frequency in the laboratory frame. We can choose to evaluate the norm of each packet in a region where $v(\tau)$ is approximately constant, as it is conserved in $\zeta$. For the three frequency modes corresponding to $\delta n=0$ (IN, P, N), the previous approximation is justified if the cavity is long compared with the duration of the perturbations. On the other hand, if we want to describe the process occurring in the system formed by the cavity and the probe pulse, the best approximation that can be made is to evaluate the norm of the T packet with $\delta n=\delta n_{\text{max}}$.

It is possible to verify that the solutions of  Eq. \eqref{motion2} can be expressed in the form
\begin{equation}
	\phi_{\mathcal{W}_{0}}(\tau)=C_{\mathcal{W}_{0}}\text{e}^{i\mathcal{W}(\mathcal{W}_{0})\tau},\label{modesol}
\end{equation}
where
\begin{equation}
	\left|C_{\mathcal{W}_{0}}\right|_{\text{IN},\text{T}}=1,\quad 
	\left|C_{\mathcal{W}_{0}}\right|_{\text{P},\text{N}}=\left[v^{2}(\tau)-1 \right]^{-3/4}.\label{modesol2}
\end{equation}
These relations hold if we redefine the modes $\text{P}$ and $\text{N}$ by  only considering the first two terms of Eq. \eqref{PN}, as the last term is already included in the expression for $|C_{\mathcal{W}_{0}}|_{\text{P},\text{N}}$. On the other hand, the explicit calculation of the inner product yields the following relations 
\begin{align}
	\left(\psi_{\text{IN},\text{T}},\psi_{\text{IN},\text{T}}\right) &\simeq \pm \frac{4 \pi \epsilon_{0}c^{2}}{u \hbar}\int{\dd\mathcal{W}_{0}|G_{\mathcal{W}_{0}}|^{2}}, \label{Norm1}\\
	\left(\psi_{\text{P},\text{N}},\psi_{\text{P},\text{N}}\right) &\simeq \pm \frac{4 \pi \epsilon_0c^2\Omega_0}{u \hbar} \int{\dd \mathcal{W}_{0}\frac{|G_{\mathcal{W}_{0}}|^{2}}{\mathcal{W}_{0}}}. \label{Norm2}
\end{align}

Having obtained these approximated expressions for the norms of the packets, we can give a qualitative explanation of the amplification of the analogue of Hawking radiation in the OBHL. Let us consider the evolution forwards in $\zeta$ of an initial $\text{IN}$ packet inside the cavity, as studied in Section \ref{propa} and shown in Fig. \ref{evolution} (right). Due to norm conservation, the following relations hold
\begin{equation}
	(\text{IN}_{j},\text{IN}_{j})=(\text{IN}_{j+1},\text{IN}_{j+1})+(\text{T}_{j+1},\text{T}_{j+1}), \qquad j \geq 0, \label{NormConserv}
\end{equation}
where $\text{IN}_{0}$ is just the initial $\text{IN}$ packet mentioned before. According to Eqs. \eqref{Norm1} and \eqref{Norm2}, the $\text{IN}_{j}$ ($\text{T}_{j}$) packets have positive (negative) norm, therefore Eq. \eqref{NormConserv} indicates that $(\text{IN}_{j+1},\text{IN}_{j+1})> (\text{IN}_{j},\text{IN}_{j})$, where $j\geq 0$. For each cycle of mode conversion processes at the horizons, the norm of the $\text{IN}_{j}$ packets grows by a fixed multiple. By rearranging Eq. \eqref{NormConserv} in order to express the norm of the $\text{T}_{j}$ packets in terms of the $\text{IN}_{j}$ ones, it can be seen that $(\text{T}_{j+1},\text{T}_{j+1})>(\text{T}_{j},\text{T}_{j})$, and moreover, that the norm of the $\text{T}_{j}$ packets grows (in magnitude) by the same multiple as the $\text{IN}_{j}$ packets. This results in an increase of emitted radiation when viewed forwards in $\zeta$, because the number of particles emitted is related to the expectation value of the number operator defined in terms of the T packets, as we will see in Section \ref{partcre}.

On the other hand, if the evolution backwards in $\zeta$ of the outgoing $\text{T}$ packet with negative norm of Fig. \ref{evolution} (left) is considered, then the norm conservation yields the same relation as in Eq. \eqref{NormConserv}, so that the norm of both packets grows backwards in $\zeta$, but when viewed forwards in $\zeta$, this fact is consistent with the conclusion previously stated that the Hawking process is self-amplifying \cite{Corley1999a}.


\subsection{Evolution formulas}
\begin{figure}
	\centering
	\includegraphics[height=53mm]{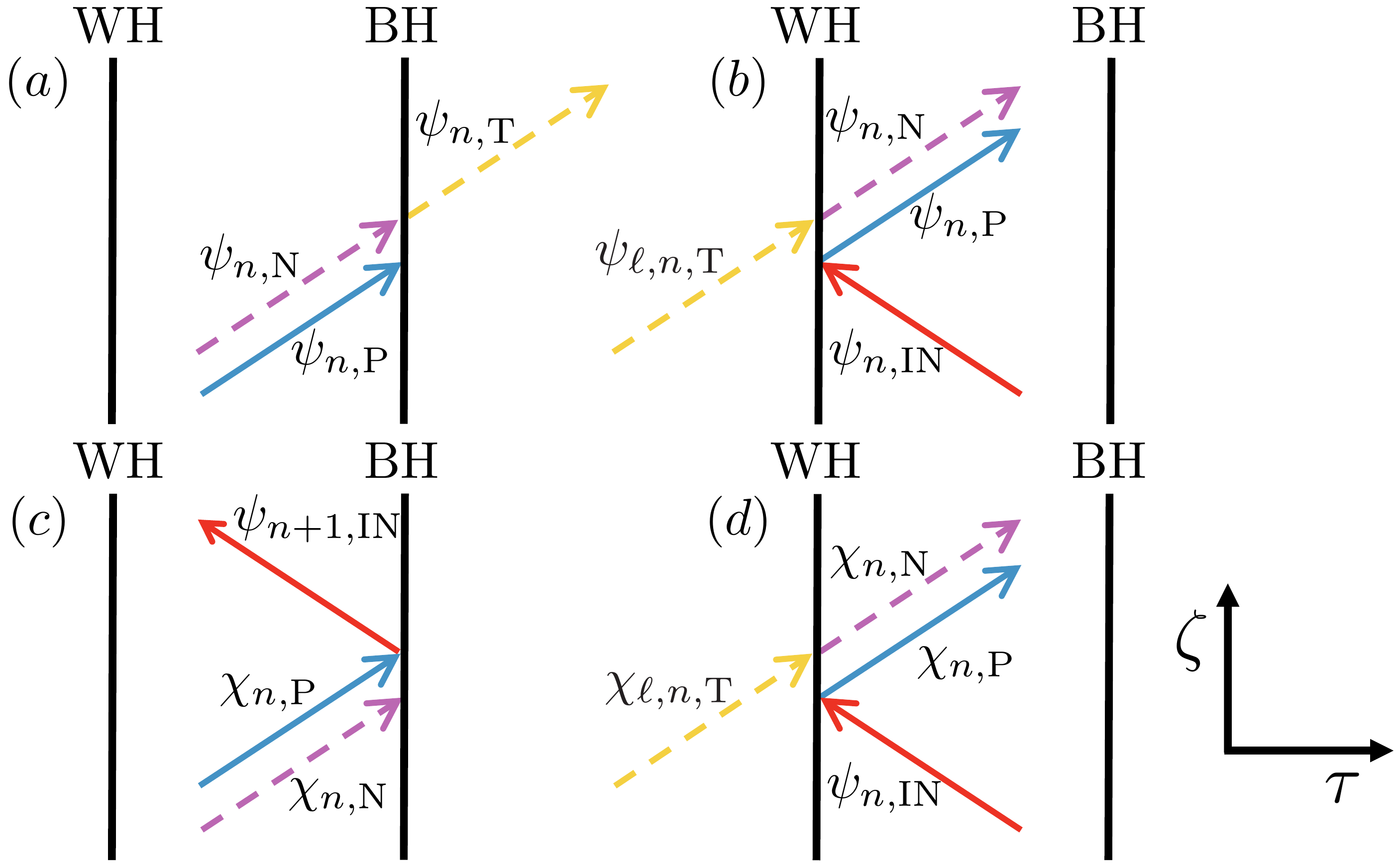}\hspace{0mm}
	\includegraphics[height=53mm]{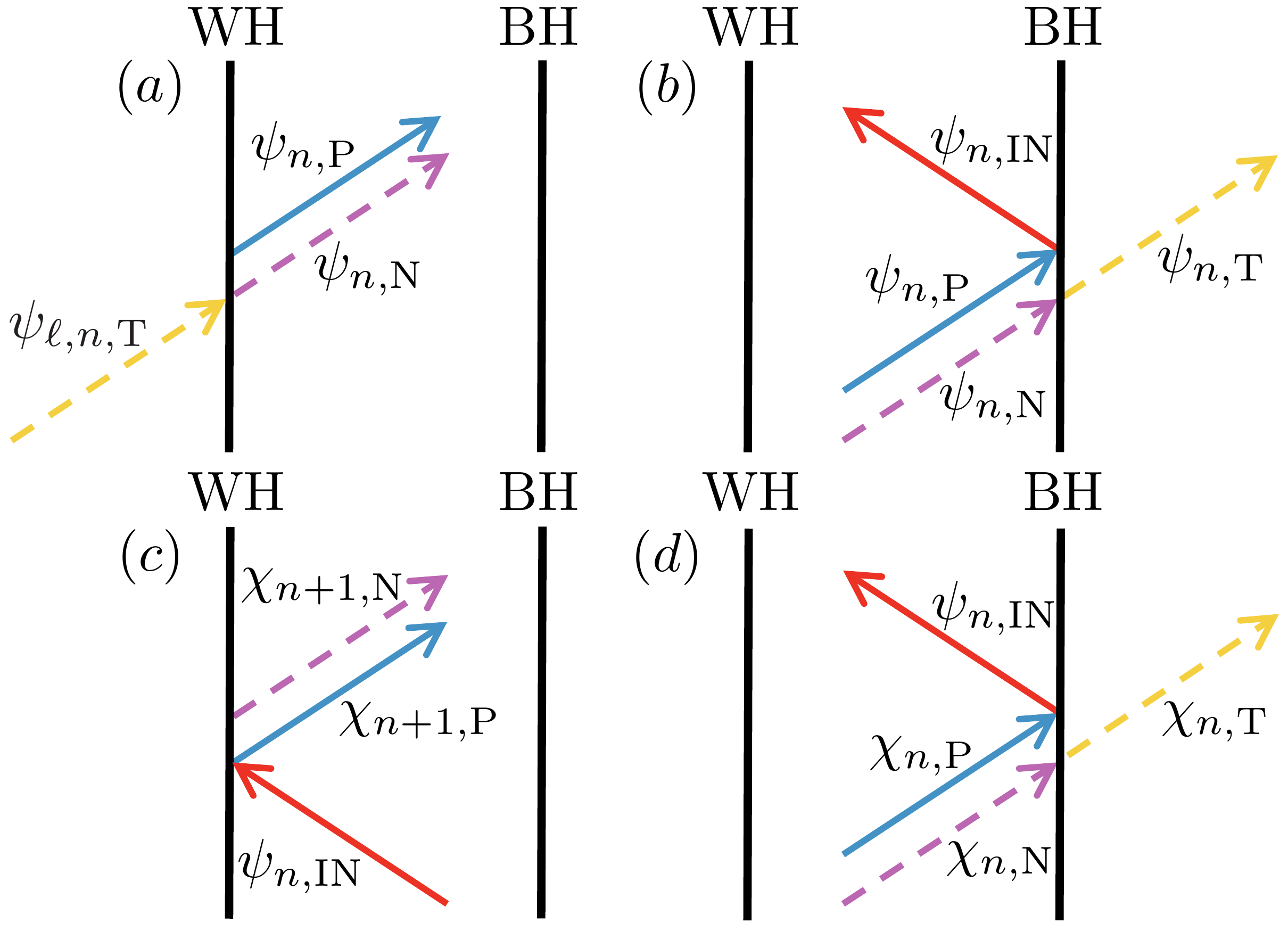}
	\caption{(Color online). Spacetime sketches of the local wave packet evolutions in $\zeta$. Evolution of an outgoing T packet backwards in $\zeta$ (left) and an incoming $\text{T}_\ell$ forwards in $\zeta$ (right). An incoming IN packet forwards in $\zeta$ requires just $(c)$ and $(d)$ from the right-hand side.}
	\label{forwardformulas}
\end{figure}

In this section we will derive the evolution formulas that describe quantitatively the propagation of modes as we outlined in Section \ref{propa}. First, we check the results (see Ref. \cite{Corley1999a}) for the evolution backwards in $\zeta$ of a final outgoing packet T, as shown in Fig. \ref{evolution} (left). The backward evolution formulas are based on the $(a)$, $(b)$, $(c)$, and $(d)$ diagrams of Fig. \ref{forwardformulas} (left). The final evolution formula describes the evolution of a pair of T packets emitted from the BH at cycles $n$ and $m$ of amplification, where $n>m$, in terms of all the packets inside the WH involved in the process. The following expression is obtained
\begin{equation}
	\psi_{\ell,n,\text{T}}+\sum_{j=1}^{n-m}{\chi_{\ell,n-j,\text{T}}}-\psi_{\ell,m,\text{T}}\longrightarrow \psi_{n,\text{T}}-\psi_{m,\text{T}}, \label{back}
\end{equation}
where the difference between the packets at the left of the $\text{WH}$, labeled by the index $\ell$, is that the $\chi_{\ell,\text{T}}$ packets account for the intermediate mode conversion processes in which there are no outgoing $\text{T}$ packets considered, whereas the $\psi_{\ell,\text{T}}$ packets result from the backward evolution in $\zeta$ of the pair of $\text{P}$ and $\text{N}$ packets that come from the final outgoing transmitted packets.

On the other hand, we now derive formulas to describe the evolution of an initial T packet that approaches the WH forwards in $\zeta$ based on the $(a)$, $(b)$, $(c)$, and $(d)$ diagrams in Fig. \ref{forwardformulas} (right). As in the previous case, the idea of the evolution formula is to describe a global solution of the equation of motion. Following all the forward diagrams, we obtain an explicit relation (see Appendix) of the form
\begin{equation}
	\psi_{\ell,n,\text{T}}\longrightarrow \psi_{n,\text{T}}+\sum_{j=1}^{m-n}{\chi_{n+j,\text{T}}}+\psi_{m,\text{IN}}.\label{preforw}
\end{equation}
If we want to see the amplification between the $n$ and $m$ cycles, we can use the expression
\begin{equation}
	\psi_{\ell,n,\text{T}}-\psi_{\ell,m,\text{T}} \longrightarrow \psi_{n,\text{T}}+\sum_{j=1}^{m-n}{\chi_{n+j,\text{T}}}-\psi_{m,\text{T}}, \label{forw}
\end{equation}
which is obtained with Eq. \eqref{preforw} (see Appendix). This is an evolution formula that describes the process where a pair of $\psi_{\ell,k,\text{T}}$ packets are sent through the WH, getting ($m-n$) $\chi_{k,\text{T}}$ packets and a pair of $\psi_{k,\text{T}}$ packets outside the BH. If the initial situation corresponds to an $\text{IN}$ mode trapped in the cavity, the evolution formula in $\zeta$ of Eq. \eqref{preforw} reduces to
\begin{equation}
	\psi_{n,\text{IN}} \longrightarrow \sum_{j=1}^{m-n}{\chi_{n+j,\text{T}}}+\psi_{m,\text{IN}}, \label{forwc1}
\end{equation}
which can be seen by following the $(c)$ and $(d)$ diagrams of Fig. \ref{forwardformulas} (right).
In contrast, for the case of two initial $\text{P}$ and $\text{N}$ modes, the corresponding evolution formula is given by
\begin{equation}
	\psi_{n,\text{P}}+\psi_{n,\text{N}} \longrightarrow \psi_{n,\text{T}}+\sum_{j=1}^{m-n}{\chi_{n+j,\text{T}}}+\psi_{m,\text{IN}}, \label{forwc2}
\end{equation}
according to the $(b)$, $(c)$, and $(d)$ diagrams.

\subsection{Particle creation}\label{partcre}
Given a normalized solution $f(\tau,\zeta)$ to the wave equation, an operator $\hat{a}(f)$ can be defined in terms of a self-adjoint operator solution $\hat{\phi}$ as $\hat{a}(f)\equiv (f,\hat{\phi} )$, where $\hat{\phi}$ satisfies the canonical commutation relations. Then, $\hat{a}(f)$ behaves as a creation or annihilation operator depending on the sign of the norm of the solution. In addition, the expectation value of the number operator $\hat{N}(f)=\hat{a}^{\dagger}(f)\hat{a}(f)$ accounts for the amount of particles in the state. To find the number of particles emitted between the $n$ and $m$ cycles (where $n>m \gg 1$), from the solution given by Eq. \eqref{back} and under the ground state conditions
\begin{equation}
	\hat{a}( \hat{\psi}_{\ell,k,\text{T}} )| 0 \rangle = 0 = \hat{a}\left( \hat{\chi}_{\ell,k,\text{T}} \right)| 0 \rangle, \qquad k \geq m, \label{Cond}
\end{equation}
where we consider normalized packets $\hat{\psi}_{\ell,k,\text{T}}$ and $\hat{\chi}_{\ell,k,\text{T}}$, the expectation value $\langle 0 | \hat{N}(\hat{\psi}_{n,\text{T}}) | 0 \rangle$ has to be calculated. Using Eq. \eqref{Cond}, the following relation holds
\begin{equation}
	\langle 0 | \hat{N}(\hat{\psi}_{n,\text{T}}) | 0 \rangle = \frac{\left(\psi_{m,\text{T}},\psi_{m,\text{T}} \right)}{\left( \psi_{n,\text{T}},\psi_{n,\text{T}}\right)}\langle 0 | \hat{N}(\hat{\psi}_{m,\text{T}}) | 0 \rangle.
\end{equation}
And the norm of the T packet can be calculated from Eq. \eqref{Norm1}. As a result the number of particles emitted at $m$ and $n$ cycles after the emission of a fixed outgoing particle is given by the expression
\begin{equation}
	\frac{\langle 0 | \hat{N}(\hat{\psi}_{n,\text{T}}) | 0 \rangle}{\langle 0 | \hat{N}(\hat{\psi}_{m,\text{T}}) | 0 \rangle } \simeq \left(1+\frac{1-\cos [\theta_{\text{P}}(\mathcal{W}_0)-\theta_{\text{N}}(\mathcal{W}_0)]}{2 \sinh^{2}\left( \pi \mathcal{W}_{0}/|\kappa| \right)} \right)^{n-m}, \label{Ampl}
\end{equation}
in which the approximation that the packets are narrowly peaked around $\mathcal{W}_0$ was made. This expression provides an analytic proof for the amplification of Hawking radiation in agreement with Eq. \eqref{NormConserv}. Furthermore, it shows the variables that control the amplification process, such as the phase difference between the modes $\text{P}$ and $\text{N}$ that propagate in the direction WH$\rightarrow$BH. The number of particles emitted from the OBHL depends on $|\kappa|$, the analogue of the surface gravity, as it is the usual case for analogue systems, including the sonic black hole laser. In addition, the value of the phase difference can lead to a maximal amplification of Hawking radiation in a WKB-condition similar to that derived for the sonic case \cite{Leonhardt2007a} or, at least, to no amplification, i.e, there is never reduction of Hawking radiation.

In particular, as the denominator of the second term of Eq. \eqref{Ampl} is positive, then the condition for maximum amplification of Hawking radiation is that the numerator reaches its maximum value, that is, whenever
\begin{equation}
	\theta_{\text{P}}(\mathcal{W}_{0})-\theta_{\text{N}}(\mathcal{W}_{0})=\left(2q+1\right)\pi,\label{condphase}
\end{equation}
with $q$ an integer. Hence, an explicit expression of the phase-difference is needed in order to find the conditions for maximum amplification. We obtain now one of such expressions using a linear approximation.

\subsection{Approximation of the phase difference}\label{secphase}
The number of particles emitted between the $m$ and $n$ cycles depends on the phase difference $\theta_{\text{P}}(\mathcal{W}_{0})-\theta_{\text{N}}(\mathcal{W}_{0})$, which in turn varies according to the value of $\mathcal{W}_{0}$ and the form of $v(\tau)$.  The explicit expressions for the phases $\theta_{\text{P}}(\mathcal{W}_{0})$ and $\theta_{\text{N}}(\mathcal{W}_{0})$ are given by 
\begin{equation}
	\theta_{\text{P},\text{N}}(\mathcal{W}_{0})=-2\theta_{C_{\text{P},\text{N}}}+\int_{\tau_\text{WH}+\epsilon}^{-\epsilon}{\dd \tau \, \mathcal{W}_{\text{P},\text{N}}\left( v(\tau),\mathcal{W}_{0} \right)}, \label{phase}
\end{equation}
where the phase $\theta_{C_{\text{P},\text{N}}}$ is that of the coefficient $C_{\mathcal{W}_{0}}$ of Eq. \eqref{modesol} and the parameter $\epsilon >0$ is introduced, so the corresponding coefficients of Eq. \eqref{modesol} acquire an $\epsilon$-dependence. By matching the solutions obtained by the Laplace transform method with the connection formulas (in which the WKB solutions were used), the explicit expression of the first term of Eq. \eqref{phase} is given by
\begin{equation}
	2 \theta_{C_{\text{P},\text{N}}}= \pi\delta_{(P,N),N}\mp\frac{4}{3}\Omega_0\epsilon^{3/2}\sqrt{2|\kappa|}
	+\frac{\mathcal{W}_0}{|\kappa|}\ln\left(2 |\kappa| \epsilon \right).
\end{equation}

On the other hand, the expression for the second term depends on $v(\tau)$. The integral can be analytically  solved only for simple velocity profiles.
Therefore, we restrict our treatment to cavities that are large compared with the perturbations, such that $v(\tau)$ is well approximated for a large portion of the cavity by the constant value of $v(\tau)$ corresponding to $\delta n=0$, which will be denoted by $v_\text{max}\equiv c/(b_1 u)$. Furthermore, near both horizons we can approximate $v(\tau)$ up to first-order in $\tau$ as
\begin{align}
	v_{\text{WH}}(\tau) &\simeq 1+|\kappa|\left(\tau-\tau_\text{WH}\right),\label{vwh}\\
	v_{\text{BH}}(\tau) &\simeq 1-|\kappa|\tau. \label{vbh}
\end{align}
Then, the times $\tilde{\tau}_{\text{WH}}$ and $\tilde{\tau}_{\text{BH}}$, where the approximated functions $v_{\text{WH}}(\tau)$ and $v_{\text{BH}}(\tau)$ reach $v_\text{max}$ can be obtained as
\begin{align}
	v_\text{WH}(\tilde{\tau}_{\text{WH}})= v_{\text{max}}&
	\Rightarrow \tilde{\tau}_\text{WH}=\tau_\text{WH}+\frac{v_{\text{max}}-1}{|\kappa|}, \label{tauWH}\\
	v_{\text{BH}}\left(\tilde{\tau}_{\text{BH}}\right)=v_{\text{max}}&
	\Rightarrow \tilde{\tau}_\text{BH}=-\frac{v_{\text{max}}-1}{|\kappa|}. \label{tauBH}
\end{align}
In this form, we divide the interval $[\tau_\text{WH},0]$ into three regions, as shown in Fig. \ref{vtau}. The values of $v(\tau)$ in each region are
\begin{equation}
	v(\tau) =
	\left\{
	\!
	\begin{aligned}
		&v_{\text{WH}}(\tau) & &\text{ if } \tau_\text{WH}\leq\tau<\tilde{\tau}_{\text{WH}}\\
		&v_{\text{max}} & &\text{ if } \tilde{\tau}_{\text{WH}} \leq \tau \leq \tilde{\tau}_{\text{BH}}\\
		&v_{\text{BH}}(\tau)& &\text{ if } \tilde{\tau}_{\text{BH}}<\tau \leq 0\
	\end{aligned}
	\right.
\end{equation}

With this model for $v(\tau)$, the second term on the right-hand side of Eq. \eqref{phase} can be immediately 
evaluated for $\theta_{\text{P},\text{N}}$ as
\begin{align}
	\theta_{\text{P},\text{N}}(\mathcal{W}_{0}) \simeq &
	\left(\pm\Omega_{0}\sqrt{v_\text{max}^2-1}+\frac{\mathcal{W}_0 v_\text{max}}{1-v_\text{max}^2}\right)
	(\tilde{\tau}_\text{BH}-\tilde{\tau}_\text{WH})
	\pm\Omega_0\frac{\sqrt{8|\kappa|}}{3}\left[(\tilde{\tau}_\text{WH}-\tau_\text{WH})^{3/2}+
	(-\tilde{\tau}_\text{BH})^{3/2}\right]\nonumber\\
	&+\frac{\mathcal{W}_0}{2|\kappa|}\ln\left[\frac{\epsilon^{2}}{(\tilde{\tau}_{\text{WH}}-\tau_\text{WH})(-\tilde{\tau}_\text{BH})} \right]-\pi \delta_{(\text{P},\text{N}),\text{N}}.
\end{align}
Therefore, after substituting the values of $\tilde{\tau}_\text{WH}$ and $\tilde{\tau}_\text{BH}$ from Eqs. \eqref{tauWH} and \eqref{tauBH}, the phase difference is approximately given by
\begin{equation}
	\theta_{\text{P}}(\mathcal{W}_0)-\theta_{\text{N}}(\mathcal{W}_{0}) \simeq \frac{4\Omega_0}{|\kappa|}\left(\frac{2\sqrt{2}}{3}-\sqrt{v_\text{max}+1}\right)(v_\text{max}-1)^{3/2}-2\Omega_0\tau_\text{WH}\sqrt{v_\text{max}^2-1}+\pi. \label{phase2}
\end{equation}
This expression can be substituted in Eq. \eqref{Ampl}, so that the number of particles outside the cavity can be approximated in terms of the separation $\tau_\text{WH}$, which should be useful in experimental designs.

Under this approximation, Eq. \eqref{phase2} shows that the phase difference depends linearly on the separation $\tau_\text{WH}$ between the horizons. This agrees with the qualitative argument given in Ref. \cite{Corley1999a}. On the other hand, as both $v_\text{max}$ and $\Omega_{0}$ depend on the dispersion relation in the absence of perturbation, then $\kappa$ and $\tau_\text{WH}$ are the only parameters that depend on the properties of the cavity.

If we fix the dispersion relation and consider the phase difference as a function of $\kappa$ and $\tau_\text{WH}$, the possible values $\kappa_{\text{res}}$ and $\tau_\text{WH-res}$ that yield a maximum amplification of Hawking radiation satisfy
\begin{equation}
	\frac{2\Omega_0}{|\kappa_{\text{res}}|}\left(\frac{2\sqrt{2}}{3}-\sqrt{v_\text{max}+1}\right)(v_\text{max}-1)^{3/2}-\Omega_{0}\tau_{\text{WH-res}}\sqrt{v_{\text{max}}^{2}-1}=q \pi,
	\label{tauc}
\end{equation}
where Eq. \eqref{condphase} was used.
Given the definition of $\kappa$ in terms of the $\tau$-derivative of the additional contribution $\delta n$, it should be noted that Eq. \eqref{tauc} indicates that only certain values of the temporal duration of the cavity and the steepness of the pulses produce maximum amplification of Hawking radiation.

If instead of approximating the cavity with the linear model we just described, we use a square barrier of height $v_{\text{max}}$, the approximated expression for the phase difference, neglecting $\epsilon$-dependent terms, is given by the last two terms of Eq. \eqref{phase2} because in that case $|\kappa |\rightarrow\infty$.

\section{Propagation of pulses in an optical black hole laser}\label{num}
In order to study the propagation of a probe pulse trapped in a cavity constituted by other pulses, usually the nonlinear Schr\"odinger equation (NLSE) can be used \cite{Agrawal2013}. However, as the experiments of Hawking radiation in fibers need ultra-short pulses to achieve measurable radiation \cite{Bermudez2016a} and are usually performed in photonic crystal fibers, modifications must be made to include higher nonlinear terms and a more realistic dispersion of the fiber \cite{Amiranashvili2015}.

In particular, propagation equations for ultrashort pulses based on the slowly varying envelope approximation (SVEA) present theoretical difficulties, as the field and the envelope coexist and evolve on the same scale \cite{Amiranashvili2015}. In consequence, the electric field $E(t)$, which is a real quantity, is replaced by a complex analytic signal $\mathcal{E}(t)$ \cite{Amiranashvili2016}, such that the relation between both quantities is given by
\begin{equation}
	E(t)=\frac{\mathcal{E}(t)+\mathcal{E}^{*}(t)}{2}.
\end{equation}
In terms of the Fourier transform $\mathcal{E}_{\omega}$ of the analytic signal, the equation of propagation of the probe pulse propagating in the $z$-direction is given by
\begin{equation}
	\left[i\partial_{z}+\beta(\omega)\right]\mathcal{E}_{\omega}(z)+\gamma \frac{\omega^{2}}{c^{2}\beta(\omega)}P_{\omega}(z)=0, \label{prop}
\end{equation}
where $P$ is a polarization term, and $\gamma$ stands for the nonlinear parameter, proportional to the nonlinear index of refraction \cite{Agrawal2013}. A shift $-\omega/u$ must be included in Eq. \eqref{prop} so that it is solved in the co-moving frame, resulting in the expression
\begin{equation}
	\left[i\partial_{z}+\beta(\omega)-\frac{\omega}{u}\right]\mathcal{E}_{\omega}(z)+\gamma \frac{\omega^{2}}{c^{2}\beta(\omega)}P_{\omega}(z)=0.
\end{equation}
Following the approach of Ref. \cite{Rubino2012}, we may approximate the polarization term by the expression
\begin{equation}
	P(z,\tau)=2n_{0}|\mathcal{E}_{c}(\tau)|^{2}\mathcal{E}(z,\tau),\label{pol}
\end{equation}
where $\mathcal{E}_{c}$ stands for the analytic signal of the solitons that constitute the cavity that traps the probe pulse. Therefore, we consider the cavity as a fixed background, $\mathcal{E}_{c}\neq \mathcal{E}_{c}(z)$, which can be achieved in a regime where the probe is weaker than the solitons. In order to isolate the effects of the cavity from other nonlinear ones, from now on we approximate that the cavity stays in $\delta n_{\text{max}}$ after reaching this value from the inside. The propagation of the probe pulse through the cavity should show the existence not only of the modes corresponding to positive frequencies in the laboratory frame ($\text{IN}$ and $\text{P}$), but also of the negative ones ($\text{N}$ and $\text{T}$), as they are an essential part of the optical analogy of Hawking radiation and, of course, of the formulation of the OBHL. Furthermore, there is experimental evidence of their existence \cite{Steinhauer2014,Rubino2012prl,Biancalana2012}.

\begin{figure}
	\centering
	\includegraphics[height=70mm]{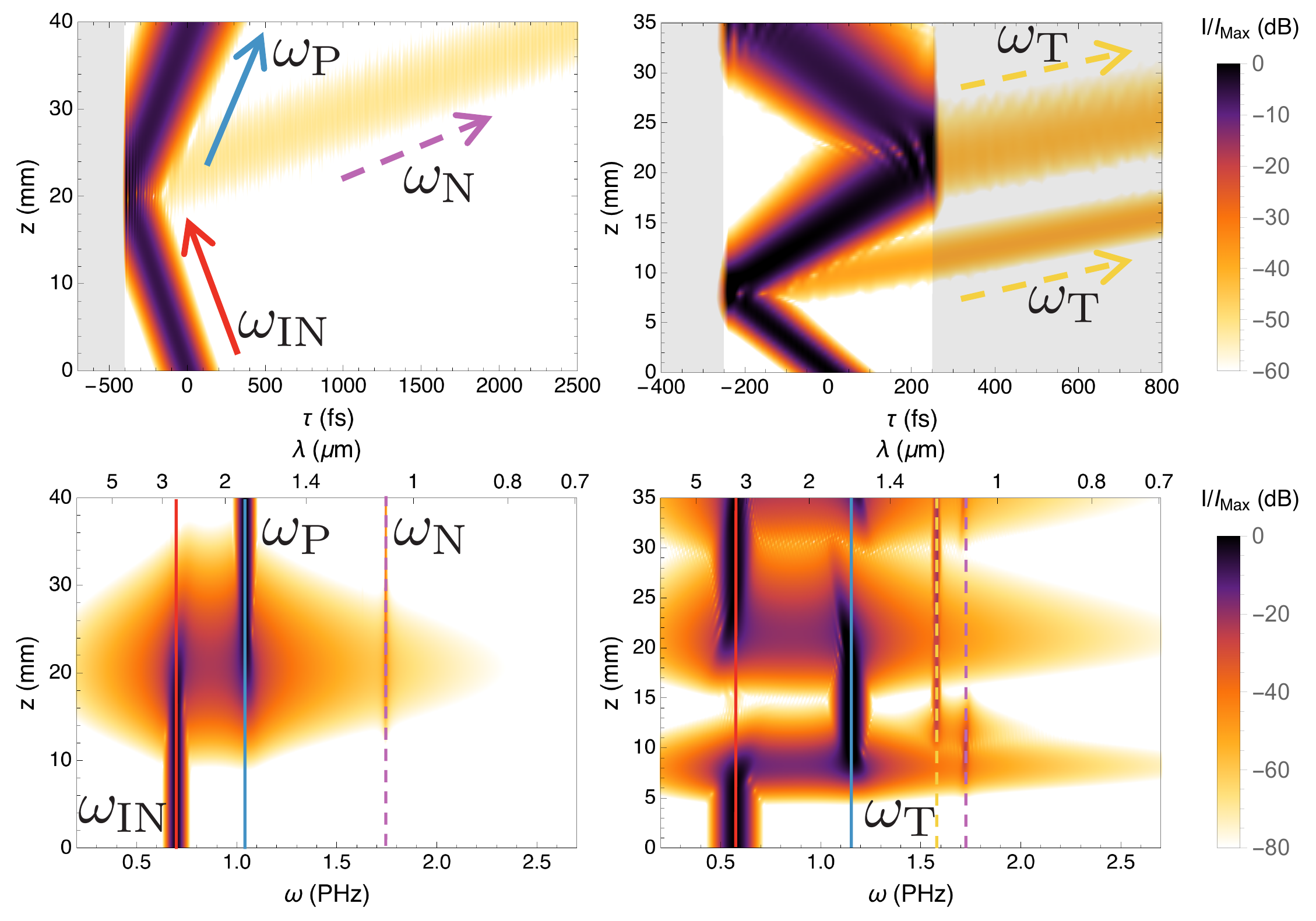}
	\caption{(Color online). Time (top) and spectral (bottom) evolution of a probe pulse of frequency $\omega_{\text{IN}}$ in the presence of a fixed background (gray). On the left-hand side there is just one soliton situated at $-400$ fs, that acts as a white hole horizon for $\omega_{\text{IN}}$ and the mode conversion process yields a pair of packets centered at $\omega_{\text{P}}$ and $\omega_{\text{N}}$. On the right-hand side, a full cavity is shown, including a signal in $\omega_{\text{T}}$ obtained in two ways, from a transformed $\omega_{\text{N}}$ and through mode conversion from $\omega_{\text{P}}$.}
	\label{timefreqev}
\end{figure}

We consider the initial form of the probe pulse as a Gaussian centered at $\omega_p$
\begin{equation}
	\mathcal{E}(0,\tau)=A_p\exp\left(\frac{\tau^2}{2\tau_p^2}\right)\exp(-i\omega_p \tau), \label{probe}
\end{equation}
where $A_p$ is the pulse amplitude and $\tau_p$ is its duration, which can be related to its full-width half maximum (FWHM). In Fig. \ref{timefreqev} (left), we show the evolution of an initial pulse centered at $\omega_\text{IN}=0.697$ PHz, of adimensional amplitude $A_p= 2 \cdot 10^{-4}$ and $\tau_{p}=60$ fs, in the presence of a soliton which produces a maximum additional contribution $\delta n_{\text{max}}=0.0023$ and a duration of $\tau_\text{sol}=2$ fs. The evolution of the probe pulse can be determined by numerically solving Eq. \eqref{prop}. The dispersion used satisfies $\omega_{h}=0.875$ PHz, and $\omega'(\omega_{h})=9.11$ THz. Two main aspects should be emphasized about the evolution of the wave packet. First, the mode conversion processes occur in the vicinity of the horizon, as expected. More importantly, after $\omega_\text{IN}$ reaches the WH it is converted into $\omega_\text{P}$ and $\omega_\text{N}$ as predicted by conservation of $\omega'_0$. It should also be noted that the packets centered at $\omega_\text{P}$ and $\omega_\text{N}$ propagate at different velocities, which is a consequence of the different values of $v_g$ and is also accounted for in our theory.

On the other hand, the evolution of a probe pulse in the presence of the whole cavity is shown in Fig. \eqref{timefreqev} (right). In this case, the cavity is formed by a pair of solitons separated in the $(\tau,\zeta)$ system by $\tau_c=500$ fs. A probe pulse of $\omega_{\text{IN}}=0.570$ PHz reaches the WH and two pulses centered at $\omega_\text{P}$ and $\omega_\text{N}$ are created as in the previous case. Then, the pair reaches the vicinity of the BH and is partially converted to $\omega_\text{T}$ and propagates outside the cavity. As in the previous case, $\omega_\text{T}$ is such that $\omega_{0}'$ is conserved. It should be mentioned that, as the trapped pulse propagates through the cavity, its temporal duration increases as a consequence of the dispersion of the fiber and, more importantly, because each frequency in the interval $(\omega_\text{min},\omega_h)$ and $(\omega_h,\omega_\text{max})$ reaches the horizon (and changes direction) at a different $\tau$.

\section{Conclusions}\label{conclusions}
We have shown that the amplification of the analogue Hawking radiation is also possible in optics for a normal dispersion regime, in contrast to the anomalous dispersion widely studied before \cite{Corley1999a,Leonhardt2007a}. In doing so, we established an equivalence between the effects of a velocity profile in a sonic context and the additional contribution $\delta n$ to the dispersion relation for the optical case, which can be seen in the rearrangement showed in Eq. \eqref{rearr} of the usual Doppler relation in optical analogues.

We describe the mode evolution inside the cavity of an OBHL for both backward and forward propagation inside the fiber. We propose an appropriate inner product for the solutions of the equation of motion for the modes and use it to verify that the amplification of the analogue Hawking radiation can be understood as a consequence of the norm conservation in the process of successive \emph{bounces} of an initial packet trapped in the cavity. In addition, we obtained an explicit expression that demonstrates the amplifying character of the phenomenon (Eq. \eqref{Ampl}). As such expression depends on the phase difference between the modes that propagate in the WH$\rightarrow$BH direction, we constructed an approximate model of the cavity and determined an explicit form of the phase difference (Eq. \eqref{phase2}). With this expression, we set a relation between the duration and the steepness of the cavity in order to obtain maximal amplification of the analogue Hawking radiation.

Moreover, by numerically solving the propagation equation for the probe pulse trapped between the perturbations, we presented its evolution as it propagates through the cavity. We observed that the mode conversion processes indeed take place in the vicinity of the horizons and that they lead to the creation of packets centered at the frequencies theoretically predicted (Fig. \ref{timefreqev}).

Although the phase difference should explicitly depend on $\omega_{0}'$, our result does not show this dependence as it is the same for both the $\text{P}$ and $\text{N}$ modes in Eq. \eqref{PN}. So it should be expected that the inclusion of additional terms in those approximations would restore an explicit dependence on $\omega_{0}'$. However, this additional terms would modify the form of the WKB solutions and thus of all the related expressions.

It should be mentioned that for values of $\delta n$ closer to the experimental ones, the frequency $\mathcal{W}_{\text{T}}$ is better approximated by evaluating the expression for $\mathcal{W}_{\text{N}}$ of Eq. \eqref{PN} with $\delta n=\delta n_{\text{max}}$. This modification would alter the form of the WKB solution and the related  expressions, in analogy with the previous case. 

Even though the simplified model of the dispersion relation (Eq. \eqref{betaapp}) reproduces the essential characteristics of light propagation in a fiber, a real dispersion relation may enable the existence of additional modes that correspond to the same value of $\omega_{0}'$. In a first approach, a more elaborated model of the dispersion relation, which  includes higher powers of $\omega$, would allow for the existence of those modes. However, it should be also taken into account that the corresponding equation of motion would necessarily have to include additional terms corresponding to the order of the powers of $\omega$, thus formidably complicating both approaches of solution.

In 2010, A. Coutant and R. Parentani presented an alternative description of the sonic black hole laser in terms of frequency eigenmodes that are asymptotically bound in space. They found that the approaches followed until then were valid under very specific conditions \cite{Coutant2010}. In that sense, the intention of our work is to provide the optical context with a more robust theoretical basis, so that the ultimate goal of observing in the laboratory the amplification of the analogue Hawking radiation is fulfilled.

It is also worth mentioning a recent work by the group of G. Agrawal \cite{Plansinis2016} where the classical trapping of a probe between two perturbations forming a cavity is used as a temporal waveguide, i.e., a waveguide that guides in time instead of in space. This system is similar to our black hole laser, and in particular, their approach of using optical modes that can be supported by the waveguide could be used for short-duration cavities, where the WKB approximation breaks down. In that case, it would be desirable to derive an expression showing the conditions of amplification in terms of the eigenmodes to establish a connection between both approaches.

On the other hand, among the research topics in analogue gravity the gauge--gravity correspondence has been recently studied \cite{Hossenfelder2015,Dey2016} for BECs. It is left to see if this result can be derived in the optical case.

\section*{Acknowledgments}
The authors would like to thank Ulf Leonhardt, Shalva Amiranashvili, and Daniele Faccio for valuable discussions. They also thank the people at the Weizmann Institute of Science (Israel) for their hospitality during part of the period in which this work was done, in particular Jonathan Drori and Yuval Rosenberg. This work was supported by RedTC 2016 and Conacyt-Mexico 280807.

\appendix
\section{Evolution formulas forwards in $\zeta$}\label{app1}
In this Appendix we will obtain the evolution formulas forwards in $\zeta$ of Eqs. (\ref{preforw}-\ref{forwc2}). Based on the sketches of Fig. \ref{forwardformulas} (right), the evolution formulas around the WH are 
\begin{alignat}{4}
	(a)&\quad &\psi_{\ell,n,\text{T}} &\longrightarrow \psi_{n,\text{P}}+\psi_{n,\text{N}}, \label{evola}\\
	(c)&\quad &\psi_{n,\text{IN}} &\longrightarrow \chi_{n+1,\text{P}}+\chi_{n+1,\text{N}},\label{evolb}
\end{alignat}
whereas about the BH they read
\begin{alignat}{4}
	(b)&\quad &\psi_{n,\text{P}}+\psi_{n,\text{N}} &\longrightarrow \psi_{n,\text{T}}+\psi_{n,\text{IN}},\label{evolc}\\
	(d)&\quad &\chi_{n,\text{P}}+\chi_{n,\text{N}} &\longrightarrow \chi_{n,\text{T}}+\psi_{n,\text{IN}}.\label{evold}
\end{alignat}
Therefore, if the solution consists of an initial $\text{T}$ packet located at the left of the WH, from Eq. \eqref{evola}, the resulting P and N packets can be evolved forwards in $\zeta$ using Eq. \eqref{evolc}, giving
\begin{equation}
	\psi_{\ell,n,\text{T}} \longrightarrow \psi_{n,\text{T}}+\psi_{n,\text{IN}}, \label{evolint}
\end{equation}
where the last mode can be evolved using Eqs. \eqref{evolb} and \eqref{evolc}. The expression of Eq. \eqref{evolint} can be evolved forwards in $\zeta$ as
\begin{equation}
	\psi_{\ell,n,\text{T}} \longrightarrow \psi_{n,\text{T}}+\sum_{j=1}^{2}{\chi_{n+j,\text{T}}}+\psi_{n+2,\text{IN}}. \label{evolint2}
\end{equation}

After $m$ cycles, the corresponding evolution formula is just a generalization of Eq. \eqref{evolint2}, that is
\begin{equation}
	\psi_{\ell,n,\text{T}} \longrightarrow \psi_{n,\text{T}}+\sum_{j=1}^{m-n}{\chi_{n+j,\text{T}}}+\psi_{m,\text{IN}},
\end{equation}
thus obtaining Eq. \eqref{preforw}. In particular, for the case $m=n$, the result obtained from Eq. \eqref{preforw} can be used to derive the formula
\begin{equation}
	\psi_{\ell,n,\text{T}}-\psi_{\ell,m,\text{T}} \longrightarrow \psi_{n,\text{T}}+\sum_{j=1}^{m-n}{\chi_{n+j,\text{T}}}-\psi_{m,\text{T}},
\end{equation}
as shown in Eq. \eqref{forw}. If the initial situation corresponds to an $\text{IN}$ packet traveling to the WH, by using Eqs. \eqref{evolb} and \eqref{evold}, Eq. \eqref{forwc1} can be obtained as a generalization of
\begin{equation}
	\psi_{n,\text{IN}} \longrightarrow \sum_{j=1}^{2}{\chi_{n+j,\text{T}}}+\psi_{n+2,\text{IN}}.
\end{equation}
In contrast, the formula shown in Eq. \eqref{forwc2} is obtained by starting the derivation shown in this Appendix with Eq. \eqref{evolc} instead of Eq. \eqref{evola}.

\bibliographystyle{ieeetr}
\bibliography{library}

\end{document}